\documentclass[letterpaper,twocolumn,10pt]{article}
\usepackage{usenix_style}

\usepackage{tikz}
\usepackage{amsmath}

\usepackage{filecontents}

\usepackage{booktabs} 

\usepackage[ruled]{algorithm2e} 

\usepackage{color}
\usepackage{verbatim, subfigure}
\usepackage{array, calc, multicol}
\usepackage{xspace}
\usepackage{enumerate}
\usepackage{graphics, epstopdf, graphicx, epsfig, psfrag, epsf, subfigure}
\usepackage{amssymb, amsfonts, amsmath, times}
\usepackage{multicol, multirow, balance, verbatim,caption}
\usepackage[mediumspace,mediumqspace,Grey,squaren]{SIunits}

\usepackage{ragged2e}

\usepackage{amsmath, amsthm, amssymb}

\usepackage{relsize}

\usepackage{url}
\usepackage{breakurl}

\makeatletter
\def\url@leostyle{%
  \@ifundefined{selectfont}{\def\UrlFont{\sf}}{\def\UrlFont{\small\ttfamily}}}
\makeatother
\usepackage{textcomp}

\renewcommand{\textrightarrow}{$\rightarrow$}

\newcommand{\eg}{{\it e.g.,}\xspace}
\newcommand{\ie}{{\it i.e.,}\xspace}
\newcommand{\etc}{{\it etc.}\xspace}

\newcommand\textvtt[1]{{\normalfont\fontfamily{cmvtt}\selectfont #1}}

\newcommand{\meddle}{\textvtt{Meddle}\xspace}

\newcommand{\haystack}{\textvtt{Lumen}\xspace}

\newcommand{\pguard}{\textvtt{Pri\-vacy\-~Guard}\xspace}
\newcommand{\recon}{\textvtt{Recon}\xspace}

\newcommand{\antShield}{\textvtt{Ant\-Shield}\xspace}
\newcommand{\antshield}{\textvtt{Ant\-Shield}\xspace}
\newcommand{\antLib}{\textvtt{Ant\-Monitor Library}\xspace}

\newcommand{\anteater}{\textvtt{Ant\-Monitor}\xspace}
\newcommand{\mturk}{\textvtt{MTurk}\xspace}

\newcommand{\zeptolab}{\textvtt{com.ze\-p\-to\-lab.ctr.ads}\xspace}

\newcommand{\taintdroid}{\textvtt{Taint\-Droid}\xspace}
\newcommand{\appfence}{\textvtt{App\-Fence}\xspace}

\newcommand{\numUsers}{220\xspace}

\newcommand\blfootnote[1]{%
  \begingroup
  \renewcommand\thefootnote{}\footnote{#1}%
  \addtocounter{footnote}{-1}%
  \endgroup
}

\begin{document}




\title{Exposures Exposed: A Measurement and User Study to \\ Assess Mobile Data Privacy in Context}

\author{
{\rm Evita Bakopoulou}\\
University of California, Irvine
\and
{\rm Anastasia Shuba\footnotemark[2]}\\
University of California, Irvine
\and
{\rm Athina Markopoulou}\\
University of California, Irvine
}

\maketitle

\blfootnote{A. Shuba was a student at the University of California, Irvine at the time the work was conducted.}%

\begin{abstract}Mobile devices have access to personal, potentially sensitive data, and there is a large number of mobile applications and third-party libraries that transmit this information over the network to remote servers (including app developer servers and third party servers). In this paper, we are interested in better understanding of not just the extent of personally identifiable information (PII) exposure, but also its {\em context} (\ie functionality of the app, destination server, encryption used, \etc) and the risk perceived by mobile users today. 
%
To that end we take two steps. First, we perform a {\em measurement study}: we collect a new dataset via manual and automatic testing and capture the exposure of 16 PII types from 400 most popular Android apps. We analyze these exposures and provide insights into the extent and patterns of mobile apps sharing PII, which can be later used for prediction and prevention. Second, we perform a {\em user study} with \numUsers participants on Amazon Mechanical Turk: we summarize the results of the measurement study in categories, present them in a realistic context, and assess users' understanding, concern, and willingness to take action.  
To the best of our knowledge, our user study is the first to collect and analyze user input in such fine granularity and on actual (not just potential or permitted) privacy exposures on mobile devices. Although many users did not initially understand the full implications of their PII being exposed, after being better informed  through the study, they became appreciative and interested in better privacy practices.
\end{abstract}

%
%

\section{Introduction}

Mobile devices  have access to a wealth of personal, potentially sensitive information and there is a growing number of applications that access, process and transmit some of this information over the network. Sometimes this is necessary for the intended operation of the applications (\eg location is needed by {\em Google Maps}) and controllable (\eg by the user  through permissions),  but for the most part, users are not in control of their data today. Applications and third party libraries  routinely transmit user data to remote servers, including application servers but also ad servers and trackers, and users have typically limited visibility and understanding of what part of their personal data is shared, with whom, and for what purpose.

With the increased interest in online privacy, there are several bodies of related work. On one hand, a number of systems have been proposed that improve data transparency and protect  personally identifiable information (PII). In general, these systems fall into three categories: (i) static analysis and application re-writing \cite{nanfinding2018, gibler2012androidleaks, egele2011pios}, (ii) dynamic analysis with a modified or rooted OS \cite{enck2014taintdroid, phonelab, vallina2013rilanalyzer, falaki2010first, wei2012profiledroid}, and (iii) VPN-based network monitoring \cite{meddle, antmonitor-poster-mobicom15, razaghpanah2015haystack, song2015privacyguard, recon15, antmonitor-arxiv, antshield-arxiv}. While these tools provide more fine-grained control over sensitive data (as opposed to just permissions), the way users engage with these privacy-preserving systems is less well studied. 
On the other hand, in the human\text{-}computer interaction (HCI) community, researchers have extensively studied how different designs for app permissions affect users' decisions on which apps to install and which permission requests are considered legitimate by users \cite{Ismail2017, Jorgensen2015, wang2015investigating, almuhimedi2015your, Lin2012}. 
 A detailed review of related work is provided in Section \ref{sec:related}.


In this paper, we are interested in understanding {\em privacy exposures}, which we define as PII transmitted by a mobile app (or third party library used by the app) on the device, over the network interface, towards a remote server. Our goal is to understand not only the extent and mechanisms of PII exposure, but also its context (\ie functionality of the app, destination server, encryption used, frequency, etc.) and the risk perceived by mobile users today. For example, location needs to be shared for  a navigation app to perform its intended and legitimate function, and should not be of concern to the user. In contrast, if the same navigation app uses a library that shares device ids with a third-party server, this is more likely a privacy {\em leak}\footnote{Most prior work \cite{recon15, vallina2018tracking, antmonitor-poster-mobicom15} refers to PII found in outgoing packets as a ``privacy leak,'' because PII is by definition private information and an outgoing packets indicates exfiltration or a ``leak.'' However, we purposely distinguish between ``privacy exposure'' (a PII contained in an outgoing packet) and ``privacy leak'' (which is an exposure that is not necessary for the intended  functionality of the app, and/or goes to a third party server, or happens in clear text). This distinction, between a PII exposure and an actual leak, can only be made based on the {\em context}, which is one of the main aspects we investigate in this paper.}
   and should be of concern to the user. We are also interested in PII actually exposed in real network traffic, as opposed to potential privacy exposures as captured by permissions. To that end, we make the following two contributions.

{\em Measurement Study.} First, we utilize a state-of-the-art mobile network monitoring tool, \anteater \cite{antmonitor-arxiv}, to intercept and inspect outgoing packets transmitted over the mobile device's network interface.  Using \anteater, we conduct two extensive and systematic experiments (one manual and one automated) on Android phones, where we test 400 most popular Android apps (as of March 2017) and we collect 47,076 outgoing packets. We identify whether these packets contain any of 16 predefined types of PII (defined in Section \ref{sec:system}), together with related information, which we collectively refer to as {\em context} including: the destination server/domain  (\ie whether it is an App Developer server or a third-party Advertisers \& Analytics server), the app category (games, shopping, navigation \etc) which reveals the intended functionality of the app, whether the PII is exposed in clear text or is encrypted, and whether the app runs in the background or foreground.   Our datasets partly confirm findings of previous measurement studies 
of mobile devices but are richer: \eg they contain previously unseen exposures over plain TCP and UDP, exposures while the app is in the background, and malicious scanning for rooted devices.  We analyze our datasets and provide insights into the extent and nature of how PII is exposed today.  We also identify 
behavioral patterns, such as communities of domains and mobile apps involved in exposing private information. These patterns can be used in the future to design automated  prediction and prevention methods. We plan to make the datasets available to the community.

{\em User Study.} Second, we perform a user study on Amazon Mechanical Turk (MTurk) with \numUsers users. We  summarize the results of the measurement study in categories, present participants with real-world scenarios of private information exposure in context (type of PII, whether it is shared with the application or a third party, use of encryption, \etc) and we ask them  to assess the legitimacy (\ie whether the information is needed for the app's functionality) and privacy risk posed. We also educate the participants on how a single piece of PII can lead to even more information being discovered when combined with data fetched from a data broker. Finally, we ask users before and after the survey what actions they would be willing to take to protect their privacy, including using free/paid privacy-enhancing tools and contributing their data to crowdsourcing. 
To the best of our knowledge, our user study is the first to collect and analyze user input in such {\em fine granularity} (\ie taking context into account) and on {\em actual} (not just potential or permitted) privacy exposures from mobile devices. We found that (i) many users did not initially understand the full implications of their PII being exposed but (ii) after being better informed through the study, they became appreciative and interested in better privacy practices. The insights gained by the study can inform the design of fine-grained data transparency and privacy preserving tools such as \anteater \cite{antmonitor-page}. 



The structure of the rest of the  paper is as follows. Section \ref{sec:related} reviews related work.
%
 Section \ref{sec:data} describes the measurement study,  including the data collection, summary and analysis. Section \ref{sec:user_study} presents the Amazon Mechanical Turk study, based on our datasets. Section \ref{sec:conclusion} concludes  the paper and outlines directions for future solutions based on the findings of this study. 
\section{Related Work} \label{sec:related}

{\bf Privacy-Preserving Tools.} There are a number of  complementary frameworks, built by different communities,  that can enhance data transparency and privacy protection on mobile devices.

{\em Permissions} are the first line of defense against unwanted access to sensitive resources. However,  they are insufficient and too coarse-grained: (i) users typically accept to install apps by default; (ii) permissions do not capture run-time behavior; (iii) they do not protect against inter-app communication and poorly documented system calls; and (iv) permissions signify {\em access} to information, which is less of a concern than {\em sharing} that information over the network.  


{\em Static analysis} and application re-writing, such as \textvtt{PiOS} \cite{egele2011pios}, \textvtt{AndroidLeaks} \cite{gibler2012androidleaks}, and \cite{nanfinding2018}, suffer from the inherent imprecision of decompilation. Furthermore, static analysis does not capture representative run-time behavior and often fails to deal with native or dynamically loaded code. 

{\em Dynamic analysis} with a modified or rooted OS include \textvtt{ProtectMyPrivacy} \cite{pmp}, \taintdroid \cite{enck2014taintdroid}, and others \cite{almuhimedi2015your, phonelab, vallina2013rilanalyzer, falaki2010first, wei2012profiledroid, Wang2017}.  Such tools are powerful not suitable for mass adoption since rooting a phone or installing a custom OS is not only a daunting task for the average user, but is also strongly discouraged by wireless providers and phone manufacturers. 

{\em VPN-based network monitoring} tools use the TUN interface to capture network packets on the device and detect whenever sensitive information is sent over the networ. Over the years, their implementation has evolved from a client-server implementation \cite{meddle, recon15, antmonitor-poster-mobicom15} to a mobile-only implementation \cite{ razaghpanah2015haystack, song2015privacyguard, antmonitor-arxiv}. State-of-the-art mobile-only network monitoring tools  include \anteater \cite{antmonitor-arxiv}, \recon \cite{recon15}, \haystack \cite{razaghpanah2016haystack}, and \pguard \cite{privacyguard}) and are amenable to crowdsourcing thanks to their implementation as a mobile-only user-space app. However, to the best of our knowledge, they have been only used  so far to collect packet traces and analyze ``privacy leaks'' found therein, not for user studies. 

{\bf Privacy ``Leak'' Datasets.} The aforementioned tools have been used before to collect datasets containing privacy ``leaks'' (see Footnote 1 for terminology) from mobile devices. Most closely related to this paper, the VPN-based monitoring tools only used so far to collect packet traces and analyze ``privacy leaks'' found therein, not for user studies.  \recon \cite{recon15} collected packet traces to train machine learning classifiers for predicting PII exposures. Razaghpanah et al. \cite{vallina2018tracking} collected cases of PII exposures from thousands of users using the \haystack app and used that data to find new advertising and tracking services. Finally, the longitudinal study of PII exposures in \cite{Ren2018} demonstrates how privacy evolves across different app versions in Android. 

{\bf User Studies.} 
Several experimental studies have analyzed mobile app behavior and several users studies have analyzed user interactions and mobile usage (\eg Mehrotra et al.  \cite{Mehrotra2017}, Tian et al. \cite{Tian2013}, \textvtt{EarlyBird} \cite{Wang2015} and Xu et al. \cite{Xu:2011:IDU:2068816.2068847}). Most closely related to this paper, are user studies that focus specifically on privacy.  

Permissions and how users interact with them have been extensively studied in \cite{Ismail2017, Jorgensen2015, wang2015investigating, almuhimedi2015your, Lin2012, Liu2016}. Almuhimedi et al. \cite{almuhimedi2015your} studied how sending users privacy nudges affected their permission settings. Wang et al. \cite{wang2015investigating} studied user decisions when presented with permission settings that are separated between apps and ad libraries. More recently, Ismail et al. \cite{Ismail2017} showed that it is possible to maintain app usability even when disallowing certain permissions. Chitkara et al. proposed a retrofitted Android system, \textvtt{ProtectMyPrivacy} \cite{pmp}, that allows users to make fewer privacy decisions by setting permissions based on third-party libraries instead of applications. 

Taking a different approach,  \textvtt{PrivacyStreams} \cite{Li:2017:PET:3139486.3130941} proposes and evaluates (with a user study) a tool for developers to write code in a more privacy-preserving way. In the web ecosystem, the work in \cite{Rao2016} studies online privacy in websites to identify mismatched user expectations and the factors that impact these mismatches. The work by Kleek et al. \cite{van2017better} is closest to our work in that they use information captured from network monitoring to see if it influences users' decisions to install apps. Unlike their work, however, we are interested in learning what users would do if given more fine-grained control over their data. 

{\bf Our work in perspective.} In this paper, we use one of the aforementioned state-of-the-art VPN-based monitoring tools, \anteater \cite{antmonitor-page, antmonitor-arxiv}, to collect and analyze packet traces and the privacy exposures found therein. This approach has the advantage that it captures actual real-world privacy exposures, as opposed to \eg potential exposures described by permissions. In addition, we compile the large volume of information from the packet traces and present it to users in a way that they can process and assess. The combination of a measurement study (volume and coverage of packet traces obtained through extensive and systematic experiments) with a user study (summarizing the information into  categories, defining context, and obtaining fine-granularity feedback from users) is one of the contributions of this paper, in addition to the detailed findings of both studies as outlined in the introduction.

\section{Measurement Study: Data Collection \& Analysis} \label{sec:data}

\subsection{Data Collection System: \antShield} \label{sec:system}

We are interested in collecting real-world cases of PII exposure, as captured in packet traces of outgoing packets (as opposed to potential exposures \eg indicated by permissions). We focus on PII that have been previously defined in related work (\eg \cite{recon15, razaghpanah2015haystack}). Specifically, we are interested in detecting PII that belong to the following list:

%
\begin{itemize}
\item Device Identifiers: IMEI, IMSI, Android ID, phone number, serial number, ICCID, MAC Address, available through Android APIs.
\item User identifiers: usernames and passwords used to login to various apps (unavailable through Android APIs); Advertiser ID, email (available through Android).
\item User demographic:  first and last name, gender, zipcode, city, \etc \xspace - unavailable through Android APIs.
\item Location:  latitude and longitude coordinates, available through Android APIs.
%
%
%
\end{itemize} 

 \begin{figure}[t!]
	\begin{center}
		\subfigure[\antShield Architecture. Data collection consists of the following steps: each packet is intercepted by \antLib, searched for any PII, and mapped to an app]{\includegraphics[width=0.99\linewidth]{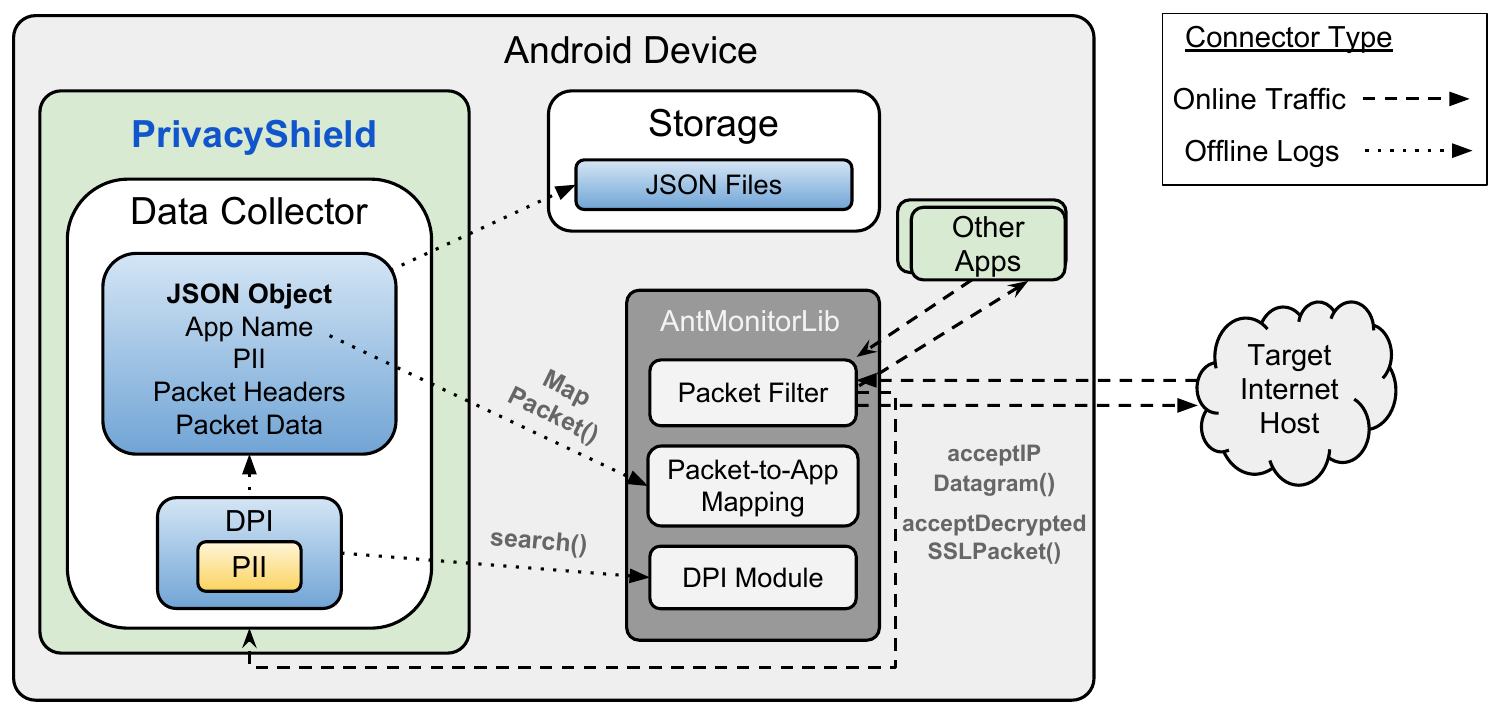}\label{fig:antShield}\hspace{10pt}}
		\subfigure[PII, including those manually entered (name)]{\hspace{6pt}\includegraphics[width=0.7\linewidth]{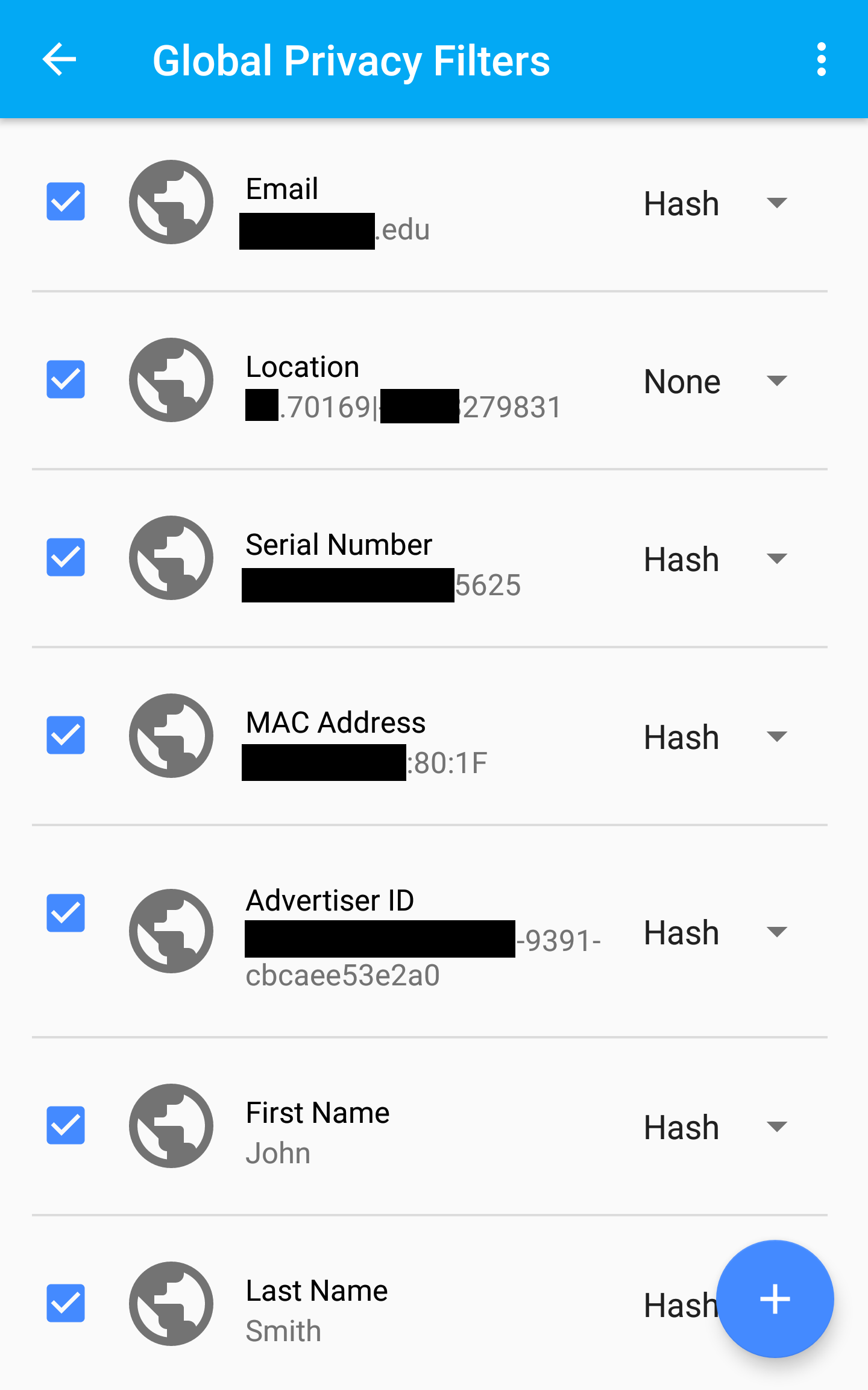}\label{fig:guiPI}}
		\caption{\antShield System used for Data Collection: Architecture and Screenshot.} 
		\label{fig:system}
	\end{center}   
\end{figure}

To collect real cases of when the PII defined above are exposed, we build on \anteater, an app that intercepts all network traffic from the device without requiring rooting. We pick \anteater since it is a representative VPN-based tool for privacy protection (see Sec. \ref{sec:related}) and it is easy to extend it for data collection. The traffic interception, along with several utility functions of \anteater have been made available as a library, and we will refer to it as the \antLib \cite{antmonitor-arxiv, antmonitor-page}. We will refer to our data collection app that extends the \antLib, as \antShield \cite{antshield-arxiv, antmonitor-page, spawc_shuba}.\footnote{The design and performance evaluation of the \antShield system is a contribution on its own. However, we consider it out of the scope of this paper. Here, we only use \antShield as a tool to collect the packet traces that are the starting point of our measurement and user studies. } As shown in Fittg. \ref{fig:antShield}, \antShield receives outgoing packets via the \textvtt{PacketFilter} interface provided by the \antLib. Note that the \antLib also implements a TLS proxy, which allows it to decrypt SSL/TLS traffic of applications that do not use certificate pinning (see \cite{antmonitor-arxiv} for details). These decrypted packets are also passed to \antShield via the \textvtt{PacketFilter} interface. Each intercepted packet (unencrypted or decrypted) is searched for PII using the \antLib's Deep Packet Inspection (DPI) module. This module implements the Aho-Corasick search algorithm to find multiple strings in one pass of a given packet. Some of the PII defined above are available to all apps via Android APIs and are thus easy to find in packets. To find PII that are unavailable through APIs, we add them to the list of strings to search for using \antshield's GUI -- see Fig. \ref{fig:guiPI}. Note that this methodology may miss PII that are obfuscated by applications prior to transmission, but as shown in \cite{continella2017obfuscation} such behavior is rare. After DPI, we use the \antLib's \textvtt{mapPacket} API call to note which app was responsible for generating the outgoing packet in question. Finally, we break the packet into any relevant fields (destination IP address/port, HTTP method, if applicable, and \etc) and save it in JSON format. Any PII found is redacted before saving the packet to allow us to share the data with the community. Any contextual information (the PII found, along with the application name, and whether or not this application was in the foreground when it generated the packet) is saved in separate JSON fields. Note that although the entire packet is not necessary for the purposes of this paper, the data could be useful later, for instance to train classifiers that predict PII exposures as in \cite{recon15} and \cite{antshield-arxiv}.

In summary, using \antShield to capture packets on the device has several advantages compared to previous datasets collected in the middle of the network: (1) we are able to accurately map each packet to the app that generated it; (2) we can keep track of foreground vs. background apps, to see what kind of data apps send while in the background; (3) we gain insight into TLS, UDP, and regular TCP traffic, in addition to HTTP and HTTPS; (4) scrubbing PII and labeling packets with the type of PII they contain is fully automated. Due to these advantages, we were able to collect comprehensive and realistic cases of PII exposures, as described in the next section.

\subsection{Datasets Collected} \label{sec:data-ours}

Using \antShield's packet capturing ability, we interacted with and collected packet traces from 400 most popular free Android apps, based on rankings in {\em AppAnnie} \cite{appannie}. We used a Nexus 6 device for our data collection, and
collected two different datasets, depending on how we interacted with apps, as described next. \footnote{These datasets are publicly available at \cite{antmonitor-page}.}

{\bf Manual Testing.}  First, in order to capture PII exposures during typical user behavior, 
we tested the top 100 apps 
in batches:  we installed 5 apps on the test device %
and then used \antShield to intercept and log packets while interacting with each app for 5min. 
After all apps in the batch were tested, we switched off the screen and waited 5min to catch  any packets  sent in the background. Next, we uninstalled each app %
and finally, turned off \antShield. 

{\bf Automatic Testing.} %
We also used the {\em UI/Application Exerciser Monkey} \cite{monkey} to automatically interact with apps.  This does not capture typical user behavior but enables extensive and stress testing of more apps. We  installed 4 batches of 100 applications each, and had {\em Monkey} perform 1,000 random actions in each tested app while \antShield logged the generated traffic. At the end of each batch, we switched off the screen of the test device and waited for 10min to catch additional exposures sent in the background. 

\begin{center}
	\begin{table}[t!]
		{\scriptsize
			\begin{tabular}[b]{|c|p{0.8cm} p{0.8cm}|} 
				\hline
				& \textbf{Auto} &  \textbf{Manual}\\[0.5ex] 
				\hline
				\# of Apps &414 & 149\\
				\hline
				\# of packets  &  21887 & 25189\\
				\hline
				\# of destination domains & 597 & 379\\
				\hline
				\# of exposures detected & 4760 &3819\\
				\hline
				\# of exposures in encrypted traffic & 1513 & 1526\\
				\hline		
				{\bf \# of background exposures} & 2289 & 639\\
				\hline
				\# of HTTP packets & 13694 &13648\\ 
				\hline
				\# of HTTPS packets &6830 &8103\\
				\hline
				{\bf \# of TCP packets} &867 &2264\\
				\hline
				{\bf \# of exposures in TCP (other ports)} & 38 &7\\
				\hline
				{\bf \# of UDP packets} & 496 &1174\\  
				\hline
				{\bf \# of exposures in UDP} & 17 &12\\ [1ex] 
				\hline
			\end{tabular}
		}
		\caption{Summary of Manual and Auto \antShield datasets collected on the device.}
		\label{tbl:summary}	
	\end{table}
\end{center}

{\bf Summary.}  Since the two (Automatic and Manual)  \antShield Datasets capture different behaviors, we describe and analyze them separately. The  \antShield datasets are summarized in Table \ref{tbl:summary}. Other state-of-the-art datasets include \recon \cite{recon15} and \cite{Ren2018}. Our datasets confirm and extend the findings of previous work, as outlined in the next section. 
Thus, in addition to being used to 
generate survey questions in our user study (Sec. \ref{sec:user_study}), our datasets provide valuable insights on their own (Sec. \ref{sec:findings}), and we will make them available to the community.

\subsection{PII Exposures Found in the Datasets} \label{sec:findings}

Our datasets provide us with insights into the current state of privacy exposures in the Android ecosystem. Some of the captured patterns were previously unknown, and are revealed for the first time here. 
For example, we were able to detect exposures happening in the background, exposures in plain TCP and UDP (not belonging to  HTTP(S) flows), 
and malicious scanning for rooted devices.

 \begin{figure}[t!]
	\begin{center}
		\subfigure[{\em Flow Free}, a puzzle game, exposes the user's city to an advertising server when the application is in the background]{\includegraphics[width=0.9\linewidth]{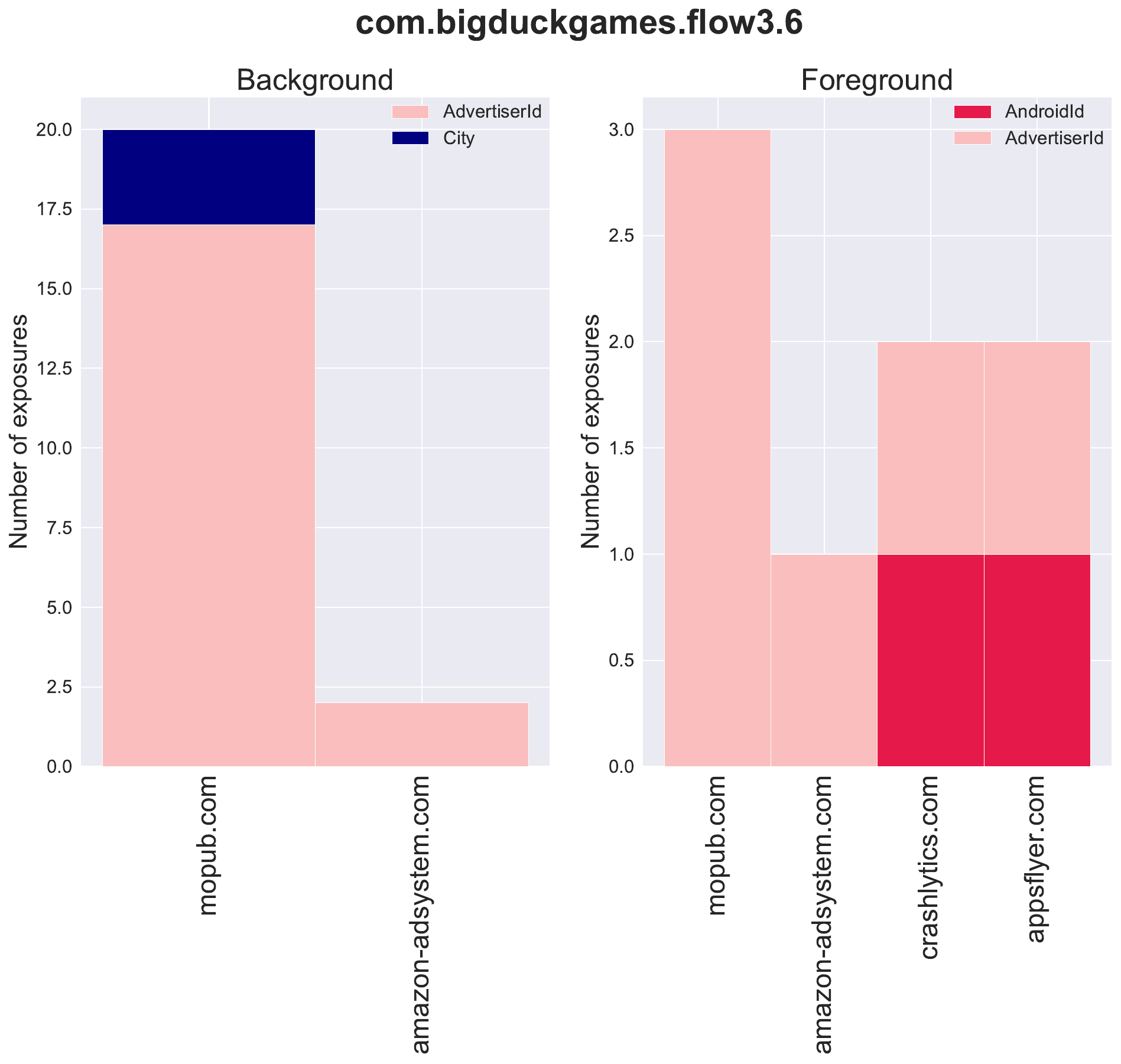}\label{fig:bigduckgames}\hspace{6pt}}
		\subfigure[{\em MeetMe}, an app for meeting people on-line contacts different domains with different PII types depending on whether it is in the background or the foreground]{\hspace{6pt}\includegraphics[width=0.9\linewidth]{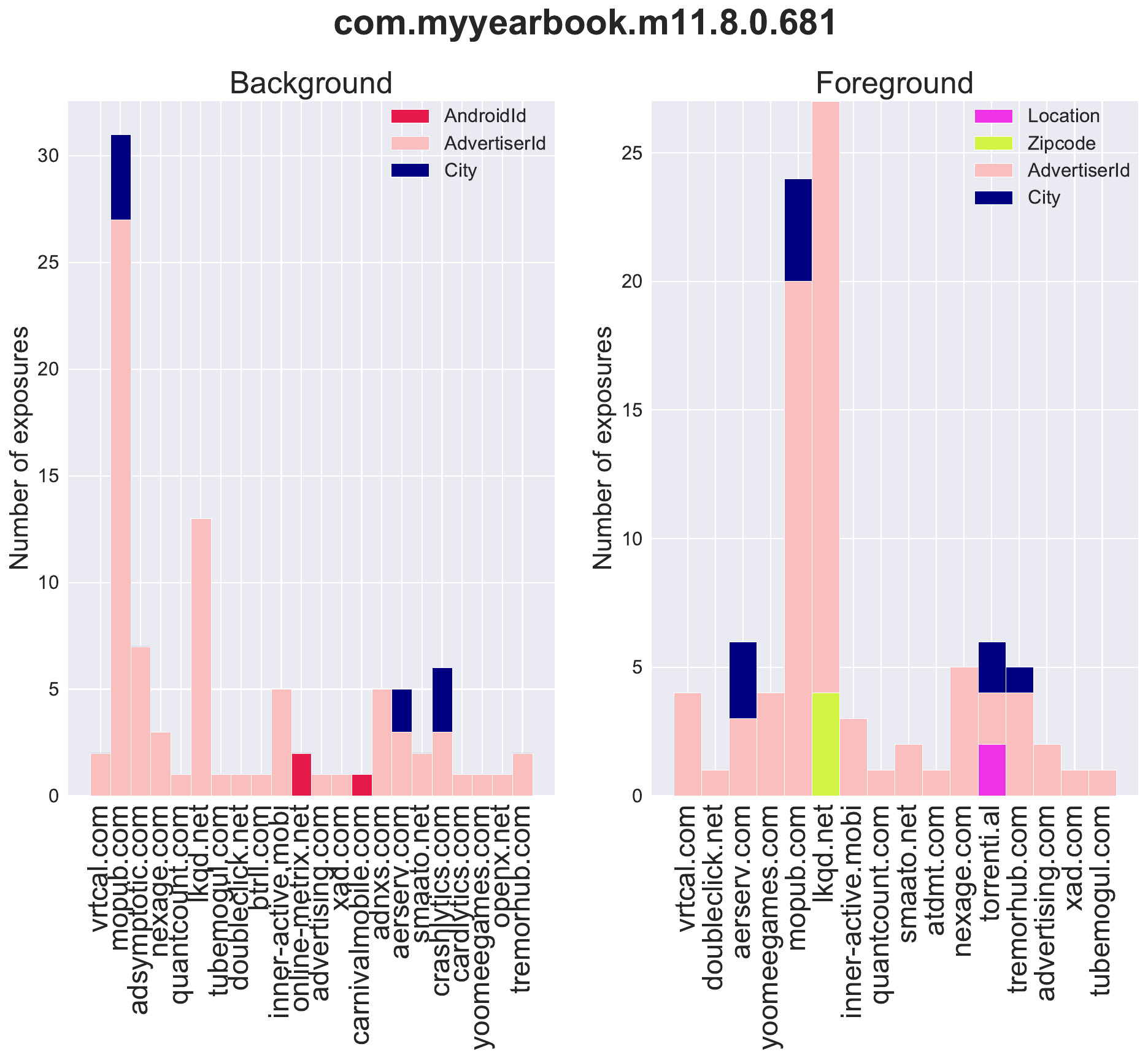}\label{fig:myyearbook}}
		\caption{Application behavior exposing PII, while running in the background vs. toreground} 
		\label{fig:bg_exposures}
	\end{center}   
\end{figure}

{\bf Background Exposures}. \antShield is in a unique position to capture exposures that happen in the background vs. foreground, and other contextual information that is only available on the device. Table \ref{tbl:summary} shows that there is a substantial number of background exposures (\eg half of all exposures in the automatic dataset) that should be brought to users' attention. 
Digging deeper, we found several interesting patterns in apps that expose PII both in the background and in the foreground. Fig. \ref{fig:bigduckgames} shows how {\em Flow Free}, a puzzle game behaves differently in the background vs. the foreground: in the foreground several device identifiers are sent to ad and analytics servers, and in the background, one of the ad servers (\textvtt{mopub.com}) also collects the user's city. Perhaps this information is needed to serve personalized ads based on the user's location, but it is unclear why it is needed when the application is in the background and no ads are being shown. Another example is {\em MeetMe}, an app for meeting people on-line, whose behavior is shown in Fig. \ref{fig:myyearbook}. In this case, the app collects less PII in the background, but is contacting more ad servers. Such findings are concerning, since apps are causing users data usage and are posing privacy risks even when the user is not interacting with the app.


\begin{table*}[t!]
	\centering
	{\small
		\begin{tabular}{|@{}l@{}|@{}l@{}|@{}l@{}|}
			\hline
			{\bf App Name} &  {\bf Leak Types}  & {\bf Port}\\[0.5ex] 
			\hline
			\xspace System  &\xspace IMEI, IMSI, AndroidId &\xspace 8080 \\
			\xspace com.jb.gosms &\xspace  AndroidId  &\xspace 10086 \\
			\xspace com.jiubang.go.music &\xspace AndroidId &\xspace 10086 \\
			\xspace air.com.hypah.io.slither &\xspace Username  &\xspace 10086 \\
			\xspace com.jb.emoji.gokeyboard &\xspace AndroidId &\xspace 10086 \\
			\xspace com.gau.go.launcherex &\xspace AndroidId  &\xspace 10086 \\
			\xspace com.steam.photoeditor &\xspace AndroidId &\xspace 10086 \\
			\xspace com.jb.zcamera &\xspace AndroidId  &\xspace 10086\\
			\xspace com.flashlight.\-brightestflashlightpro &\xspace AndroidId &\xspace 10086\\[0.5ex] 
			\hline
		\end{tabular}
		\quad 
		\begin{tabular}{|@{}l@{}|@{}l@{}|@{}l@{}|}
			\hline
			{\bf Domain Name} &  {\bf Leak Types}  & {\bf Port}\\[0.5ex] 
			\hline
			\xspace 206.191.155.105 &\xspace Username  &\xspace 454 \\
			\xspace 206.191.154.41 &\xspace Username &\xspace 454 \\
			\xspace 23.236.120.208 &\xspace AndroidId  &\xspace 10086 \\
			\xspace 3g.cn  &\xspace IMEI, IMSI, AndroidId &\xspace 8080 \\
			\xspace 23.236.120.220 &\xspace AndroidId & \xspace10086 \\[0.5ex] 
			\hline
		\end{tabular}
	}
	\caption{TCP packets (non HTTP/S) sending PII over ports other than 80, 443, 53}
	\label{tab:tcp_leaks}
\end{table*}

{\bf Non-HTTP Exposures}.  Prior state-of-the-art datasets \cite{recon15, Ren2018}  reported only HTTP(S) exposures. Table \ref{tbl:summary} reports, for the first time, exposures in non-HTTP(S), including plain TCP or UDP packets. Our dataset contains 29 UDP exposures, all of which were exposing Advertiser ID and Location. As shown in Table \ref{tab:tcp_leaks}, we also found some apps (mostly games and photo-editing apps) that exposed the Android ID over non-standard (80, 443, 53) TCP ports, such as 8080 or 10086 (a port known to be used  by trojans, Syphillis and other threats \cite{portsguide}). The destination IPs could not be resolved by DNS, indicating that the application may have hard-coded those IPs. 

\begin{table*}[t!]
	\centering
	{\scriptsize
		\begin{tabular}{|@{}l@{}|@{}l@{}|@{}l@{}|}
			\hline
			{\bf \xspace App Name} &  {\bf \xspace PII Types}  & {\bf \xspace \# Exposures \:}\\
			\hline
			\xspace com.ss.android.article.master &\xspace  \multirow{2}{*}{  \parbox{3.5cm}{City, Adid, Location, AndroidId, IMEI}}  &\xspace 752 \\
			& & \\
			\xspace com.cleanmaster.security & \xspace Adid, AndroidId  &\xspace 174\\
			\xspace com.paypal.android.p2pmobile  &\xspace \multirow{2}{*}{  \parbox{3.5cm}{ City, FirstName, LastName, Zipcode, Adid, SerialNumber, AndroidId, Password, Email}}&\xspace 131 \\
			& & \\
			& & \\
			\xspace com.offerup&\xspace \multirow{2}{*}{\parbox{3.5cm} {Adid, Username, FirstName, Location, Zipcode, AndroidId} }  &\xspace 114 \\
			& & \\
			\xspace com.cmcm.live &\xspace \multirow{2}{*}{\parbox{3.5cm}{Adid, AndroidId, Location, IMEI, SerialNumber, IMSI}}  &\xspace 114 \\
			& & \\
			\xspace me.lyft.android &\xspace  \multirow{2}{*}{\parbox{3.5cm}{City, FirstName, LastName, SerialNumber, Zipcode, PhoneNumber, Location, AndroidId}}   &\xspace 112 \\
			& & \\
			& & \\
			\xspace com.pinterest  &\xspace Adid, AndroidId &\xspace 111 \\
			\xspace com.weather.Weather  & \xspace Adid, Location &\xspace 110 \\
			\xspace com.qisiemoji.inputmethod &\xspace Adid, IMEI, AndroidId &\xspace 83 \\
	  			  
			\xspace $\cdots$ &\xspace $\cdots$ &\xspace $\cdots$ \\
			\hline
			\xspace All &\xspace All &\xspace 3039 \\
			\hline
		\end{tabular}
		\quad \quad
		\begin{tabular}{|@{}l@{}|@{}l@{}|@{}l@{}|}
			\hline
			{\bf\xspace Domain Name} &  {\bf \xspace Leak Types}  & {\bf\xspace \# Exposures \:}\\
			\hline
			\xspace mopub.com & \xspace Adid  &\xspace 2380 \\
			\xspace isnssdk & \xspace AndroidId, IMEI & \xspace805\\
			\xspace roblox.com  & \xspace Location  &\xspace 679 \\
			\xspace applovin.com & \xspace Adid  &\xspace 566 \\
			\xspace rbxcdn.com & \xspace Location  &\xspace 561 \\
			\xspace appsflyer.com  &\xspace Adid  &\xspace 549 \\
			\xspace facebook.com  & \xspace Adid  &\xspace 391 \\
			\xspace bitmango.com   & \xspace Adid  &\xspace 371 \\	
			\xspace goforandroid.com &  \xspace AndroidId &\xspace 262 \\	
			\xspace ihrhls.com  &\xspace Adid &\xspace 219 \\
			\xspace pocketgems.com  & \xspace AndroidId  &\xspace 211 \\
			\xspace ksmobile.net & \xspace SerialNumber, Location, AndroidId  &\xspace 159 \\
			\xspace tapjoy.com & \xspace Adid, AndroidId  &\xspace 151 \\
			\xspace tapjoyads.com  &\xspace IMEI, AndroidId  & \xspace147 \\
			\xspace wish.com & \xspace Adid &\xspace 139\\
			\xspace paypal.com  & \xspace AndroidId  &\xspace 131 \\		
			
			\xspace $\cdots$ & \xspace $\cdots$ & \xspace $\cdots$ \\
			\hline
			\xspace All &\xspace All &\xspace 3039 \\
			\hline
		\end{tabular}
	}
	\caption{{\small Summary of applications and domain names with HTTPS exposures in our dataset (manual and auto)}}
	\label{tab:leaks_https}
\end{table*}

{\bf HTTPS Exposures}.
Since the usage of encryption is increasing, we also collected and analyzed PII sent over HTTPS.
Table \ref{tab:leaks_https} summarizes the exposures we discovered in HTTPS traffic. The top app \textvtt{com.ss.android.\-article.\-master} is a news app, thus it makes sense for it to query the user's city, perhaps to fetch localized news. However, it is unclear why the app needs the user's IMEI (when it already has the Advertiser ID) and the specific longitude and latitude coordinates of the user. In this case, the city is needed by the app, but the IMEI and location coordinates are potentially privacy exposures. Another example is \textvtt{com.cmcm.live} - it exposes 5 different device identifiers for no apparent reason. Hence, although well-behaving apps should use HTTPS, they should also be inspected for potential privacy exposures as not all information that they gather is necessary for their functionality. We also found that the majority of top domains receiving PII over HTTPS were ad-related. Although it is expected for ad domains to receive the Advertiser ID, other PII should not be collected. These findings motivated us to conduct the user study in Sec. \ref{sec:user_study} to crowdsource answers to the question of when a privacy exposure becomes a privacy leak.

\begin{table}[t!]
	\centering
	{\scriptsize
		\begin{tabular}{|p{2.5cm}|p{2cm}|p{2.2cm}|}
			\hline
			{\bf\xspace App Name} &  {\bf\xspace Domain}  & {\bf\xspace PII Types}\\
			\hline
			\xspace com.bitstrips.imoji 10.2.32, 10.3.76 &\xspace pushwoosh.com & AndroidId \\
			\xspace com.nianticlabs\-.pokemongo 0.57.4 &\xspace upsight-api.com  & Location, AndroidId \\
			\xspace com.psafe.msuite 3.11.6 , 3.11.8  & \xspace upsight-api.com & AndroidId \\
			\xspace com.yelp.android 9.5.1 &\xspace bugsnag.com  & AndroidId \\
			\xspace com.zeptolab.ctr.ads 2.8.0  &\xspace onesignal.com & AndroidId \\
			\xspace com.namcobandai\-games.pacmantournaments 6.3.0  &\xspace  namcowireless.com & AndroidId \\
			\xspace com.huuuge.casino.slots 2.3.185 &\xspace upsight-api.com & AndroidId \\
			\xspace com.cmplay.dancingline 1.1.1  &\xspace pushwoosh.com  & AndroidId\\
			\hline
		\end{tabular}
	}
	\caption{Applications with "jailbroken" field}
	\label{tab:jaibroken_apps}
\end{table}

{\bf Checking for Rooted Devices}. We noticed a suspicious flag  called ``jailbroken'' or ``device.jail\-broken'' exposed by several apps (\eg com.bitstrips.imoji, com.yelp.android, \zeptolab, etc). This flag was found in the URI content or in the body of a POST method  in the packets, and it was set to  1 if the device was rooted, or to 0 otherwise. In Table \ref{tab:jaibroken_apps}, we show the applications that contain this field in our dataset and the domain to which the ``jailbroken'' flag is being sent. We also show other types of exposures that the particular domain collects. From the table, we see that the flag is usually accompanied with a device identifier. Several apps send this flag to the same domain (\textvtt{upsight-api.com},  an ad network), which indicates that an ad library is probably exposing this information, rather than the app itself.

\begin{figure*}[t!]
	\begin{center}
		\centering
		\includegraphics[width=0.99\textwidth]{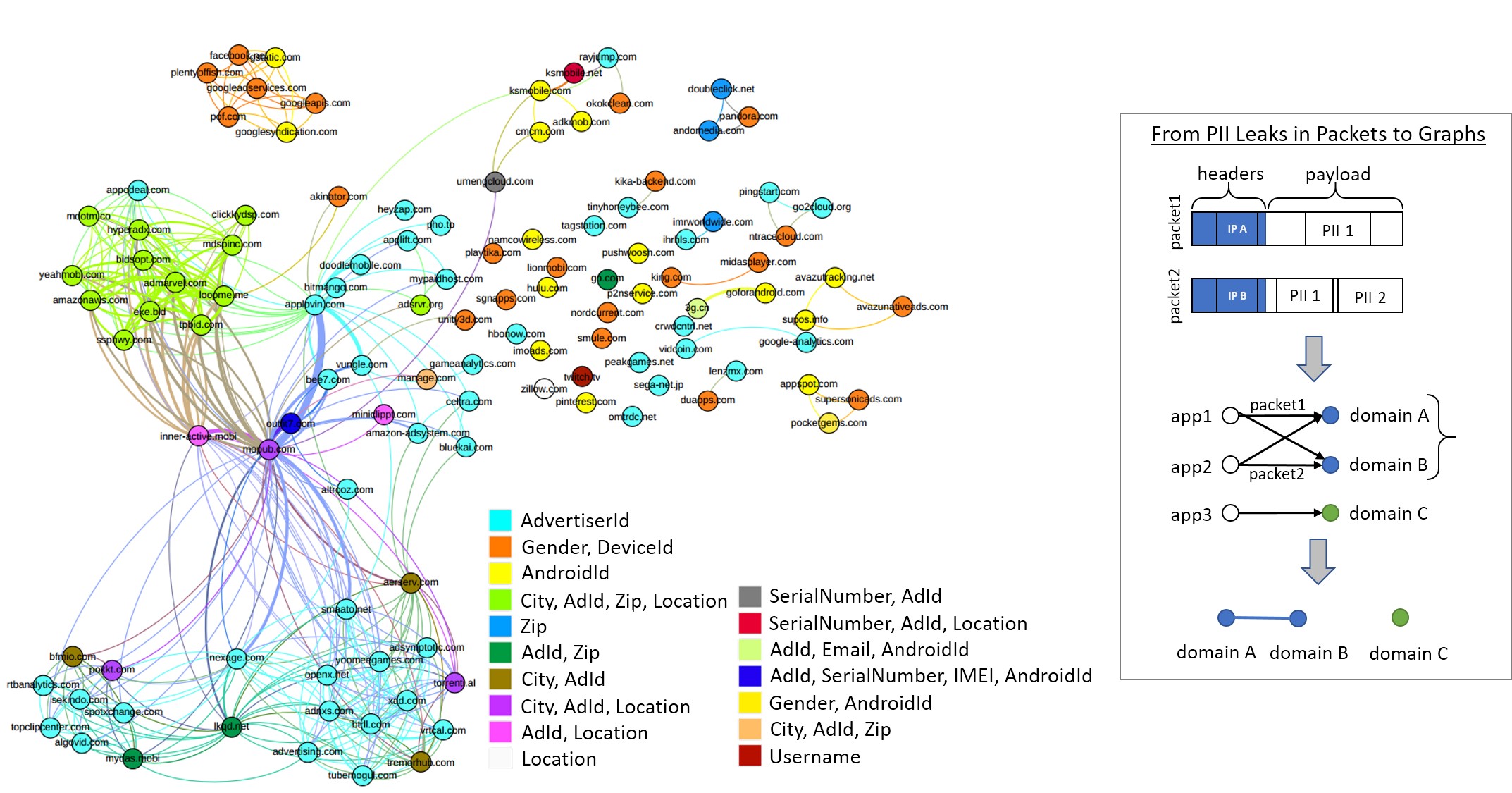}
		\caption{Understanding the behavior of app that expose PII through graph analysis of the \antShield dataset. The graph consists of nodes corresponding to destination domains and edges  representing the similarity of two domains. Two domains are similar if there are common apps that send packets with PII exposures to both domains; the more common apps  expose PII to these domains, the more similar they are, the larger the width of an edge between them. The color of a domain node indicates the types of  PII it receives. One can observe from the graph structure that domains form communities that capture interesting patterns: (1) The large communities on the left and bottom consist mostly of ad networks; ad exchanges are nodes in between ad communities; (2) Facebook/Google domains are a different community on their own, on the top left; (3)  small apps contact only their own domain, leading to isolate domain nodes; (4) domains in the same community receive the same type of PII (as indicated by the color of nodes).}
		\label{fig:graph}
		\vspace{-15pt}
	\end{center}   
\end{figure*}

{\bf Behavioral Analysis of PII Exposures.}  An interesting direction for analyzing the \antShield dataset is via behavioral analysis. For instance, we can ask: (i) what can the communication between mobile apps and destination domains reveal about tracking and advertising? (ii) what type of information exposes to what domains and how to define similarity of apps or domains with respect to exposures? Fig. \ref{fig:graph}\footnote{This figure was first presented in Fig. 3 of \cite{antshield-arxiv} and is repeated here for completeness.} showcases one graph that visualizes similar destination domains with respect to exposures they received, as captured in the \antShield dataset.
We define  two domains to be similar if they are contacted by the same set of applications (see the box on the right inside Fig. \ref{fig:graph}).
For example, domains A and B are similar because they are contacted by two apps (app1, app2).
We depict the similarity of domains A and B as an edge on the graph of domains, at the bottom of the box.
This data can be readily extracted from our trace, together with the type of information that was transmitted from apps to domains.

The graph depicted on the left side of Fig. \ref{fig:graph} shows a projection of the underlying bipartite graph (middle step in the box) on domains (last step in the box); the graph is plotted  and analyzed using Gephi \cite{bastian2009gephi}. Nodes in this graph represent domains; the edges indicate similar nodes as per above definition; the width of the edge indicates the number of common applications; and the domain color corresponds to the type of exposed PII. The clusters of domains in the graph are the output of a community detection algorithm, which is a heuristic that tries to optimize modularity.\footnote{The main idea is that for specific node $i$, it tries to assign different communities of its neighbors like node $j$'s community as $i$'s community and compute the gain of modularity for whole network. The community which maximize the modularity will be the proper one. If the gain of modularity be negative or zero, $i$ keeps its community. This process is an iterative process which is done for all nodes. This algorithm is implemented in Gephi software \cite{bastian2009gephi}, and works with weighted graphs also.}

%
%
Interesting patterns are revealed in Fig. \ref{fig:graph}. First, advertising is the result of coordinated behavior.
For example, it is easy to identify ad exchanges: \textvtt{mopub.com} is in the center of all communication; and \textvtt{inner-active.mobi} and \textvtt{nexage.com} are also clearly shown as hubs.
All three large communities on the bottom and left of the graph correspond to ad networks.
Second, on the top left, there is a community of domains that belong mostly to Google and Facebook, and two domains (\textvtt{pof.com} and \textvtt{plentyoffish.com}) of a dating service.
The latter could be because the dating app also sends statistics (\eg for advertising purposes) to Google and Facebook, in addition to its own servers, as suggested by the type of PII being sent (gender and device ID, represented by the yellow color).
Third, not all domains belong to a community: some are well-behaved and are contacted only by their own app.
For instance the white-colored domain \textvtt{zillow.com} towards the bottom center of the graph is an isolate node and only receives information about the user's location, which makes sense since it provides a real-estate service.
Another example is the blue-colored domain \textvtt{hbonow.com}: it is only contacted by its own app and only receives the Advertiser ID to serve ads.
Another observation from Figure \ref{fig:graph} is that most domains in the same community receive the same type of PII (as indicated by the domain color).
This can be explained by the common ad libraries shared among different apps that fetch the same PII.

In general, similarity of apps and domains based on their PII exposure found in their network activity can be exploited to detect and prevent abusive behavior (\eg advertising, tracking, or malware) in mobile traffic. This is one promising direction for future work.
%


\section{User Study: Mobile Data Privacy in Context} \label{sec:user_study}

In this section, we design a user study on Amazon Mechanical Turk (MTurk) in order to assess user's awareness and understanding of mobile data exposure, as well as their level of concern and potential for adopting solutions. We use the datasets collected in the measurement study in the previous section to present participants with real-world scenarios of private information that was actually exposed by mobile apps in our experiments. We present the user with information about the types of PII exposed,  as we as information about the {\em context} this exposure occured, \ie whether the PII  is shared with the application or a third party server;  what was the app category/intended functionality; whether it is shared in clear or encrypted text, etc.   We then ask the users to assess the legitimacy (\ie  whether the information is needed for the app's functionality) and the risk posed by the particular PII type exposed in that particular context. 
We also educate the participants on how a single piece of PII can lead to even more information being discovered when combined with data fetched from a data broker. Finally, we ask users before and after the survey whether they would use privacy enhancing tools. To the best of our knowledge, this user study is the first one that collects and analyzes user input in such fine granularity (context)  and on actual (not just potential or permitted) privacy exposures at large scale (as found in the packet traces of the measurement study).
%
%
Section \ref{sec:study_design} presents the design rationale and questions asked in the MTurk study.  Section \ref{sec:study_results} summarizes and analyzes the responses from \numUsers participants. 

\subsection{User Study Design} \label{sec:study_design}

{\bf MTurk Setup.\footnote{We went through the IRB process in our institution and obtained exempt research registration HS\# 2018-4272  ``Amazon Mechanical Turk Survey on Mobile Data Privacy''}.} We designed a Human Intelligence Task (HIT) on Amazon Mechanical Turk (MTurk) \cite{mturk}
and restricted it to workers who are based in the U.S., are at least 18 years old, have completed at least High School (in the US), and own a smartphone or a tablet device. 
Our study was approved by the Institutional Review Board (IRB) of our Institution (details are omitted from this double-blind version). 
The workers were rewarded at a rate of \$0.10 per minute of their time -- a standard followed in other studies \cite{Lin2012, Ismail2017, Mason2010}. We allotted 30 minutes for the completion of our HIT, but the majority of workers completed it within 13 minutes approximately. The participants had to pass at least one of three attention check questions in order to have their HIT approved and to receive the \$3.00 payment.
 The HIT was open for 9 days in early May 2018. 
 At the end, we analyzed the responses of \numUsers workers that passed the attention test.

{\bf Demographic questions.} First, we asked a set of demographic questions, such as educational level, age, and employment sector (tech vs. non-tech). We also asked what kind of mobile OS they use and how many different apps they use daily.  In addition to these questions, we added three attention-check questions to prevent workers from gaming the system by providing answers randomly. We discard answers from participants that failed to correctly answer all three attention check questions. 

\begin{figure*}[t!]
	\begin{center}
		\subfigure[Definitions used to categorize a PII exposure.]{\includegraphics[width=0.85\textwidth]{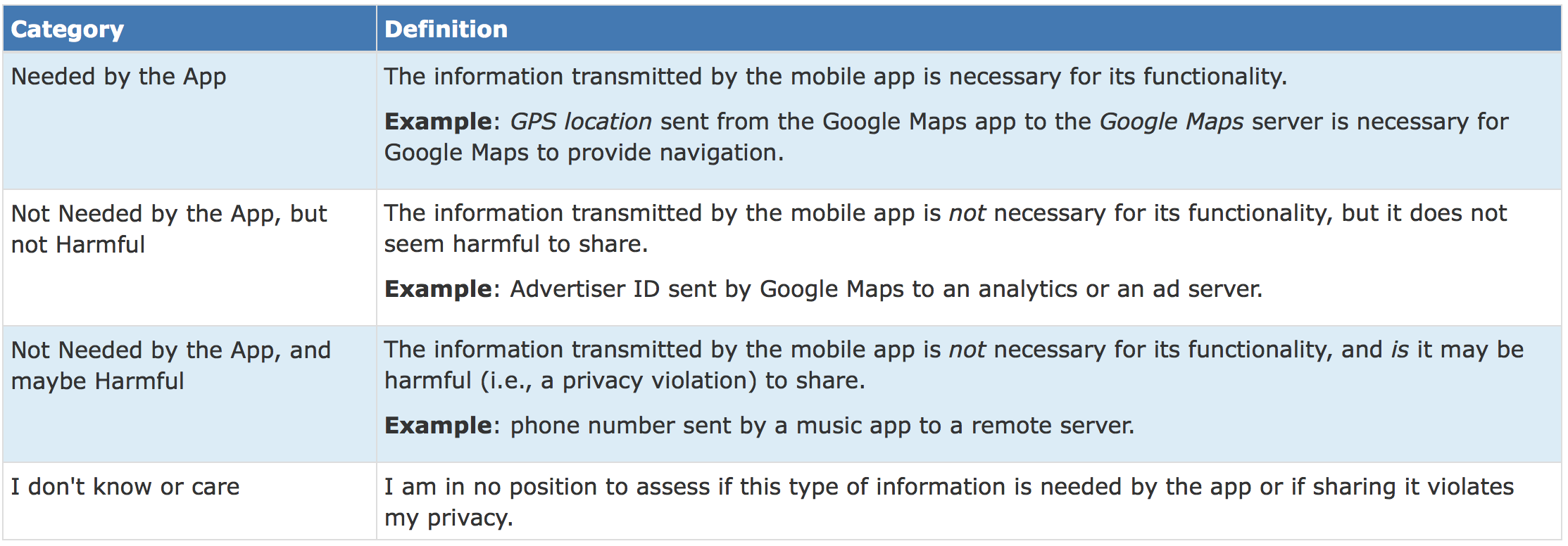}\label{fig:survey_table_categories}}
		
		\subfigure[Definitions used to describe additional context of a PII exposure.]{\includegraphics[width=0.85\textwidth]{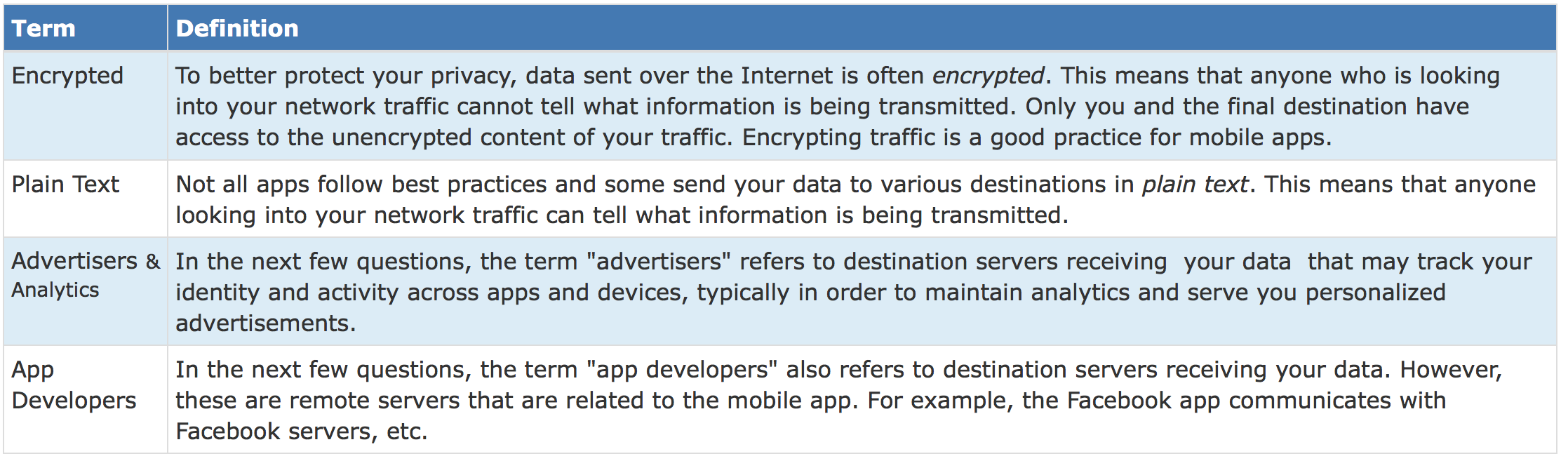}\label{fig:survey_table_terms}}
		
		\caption{Terms defined before being used in the categorization tasks.} 
		\label{fig:informative_tables}
	\end{center}   
\end{figure*}

{\bf Categorization questions.} The main goal of our study is to learn how concerned are users about privacy exposures in different contexts, defined as: the type of PII exposed, the app category, whether the information was shared with a relevant application server (thus may be useful for the functionality of the app) or third party advertiser and analytics servers, and whether it is shared in plain text or is encrypted. %
To that end, we first defined these terms and categories as shown in Fig. \ref{fig:informative_tables}.  
%

\begin{figure*}[t!]
	\begin{center}

		\includegraphics[width=0.8\linewidth]{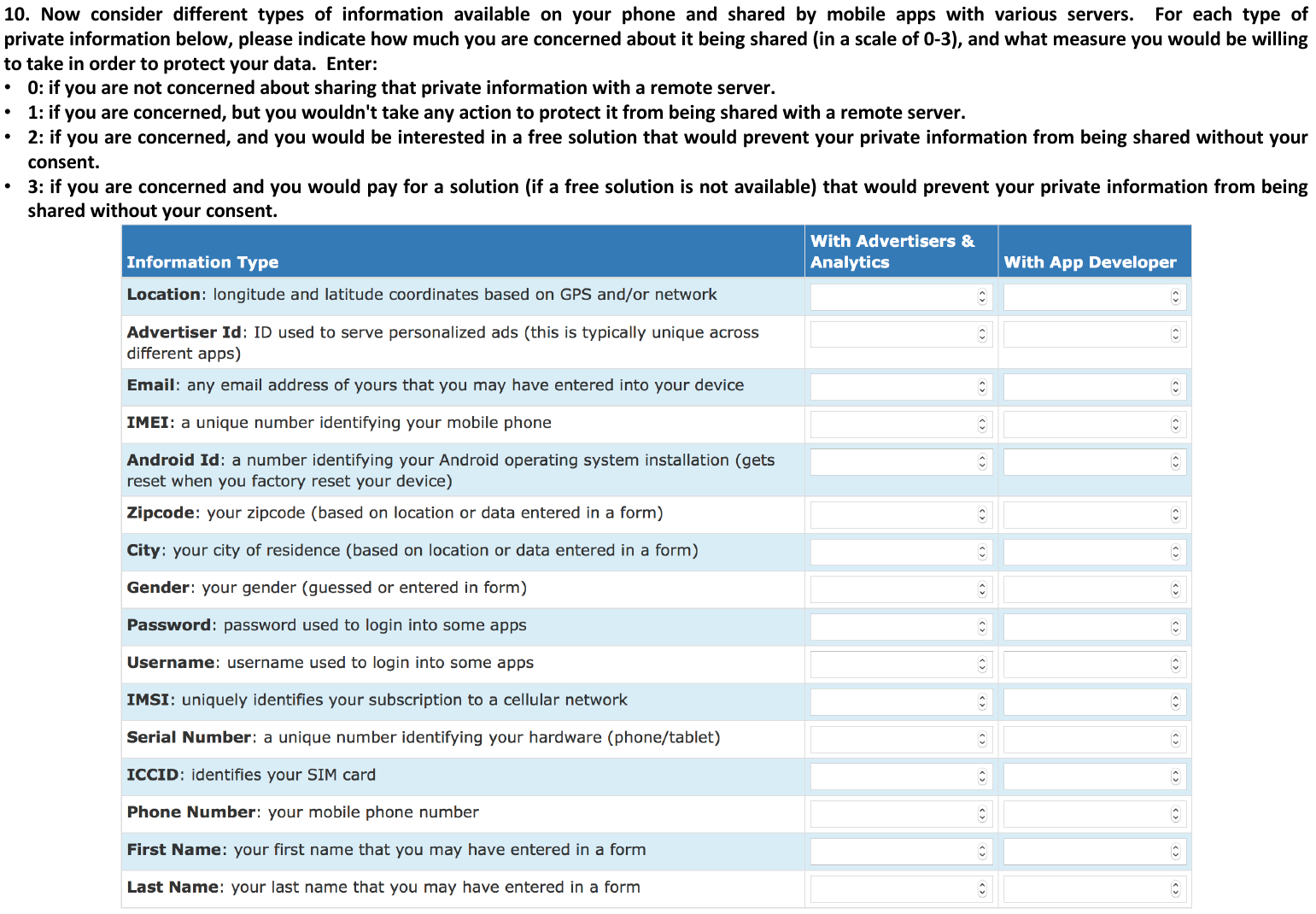}\hspace{6pt}
		%
		\caption{Task to assess how comfortable users are with sharing certain {\bf PII type} with certain type of {\bf remote server}.} 
		\label{fig:question_10}
	\end{center}   
\end{figure*}

First, we asked the participants how comfortable they are with sharing various {\bf PII types} with different types of {\bf remote servers}, as depicted in Fig. \ref{fig:question_10}. Different PII types include: various device ids (such as phone number, IMEI, IMSI, ICCID, Android Id, etc), user ids (\eg email, Advertiser Id, username and password), location (GPS coordinates), and demographic information (\eg gender, city, zipcode, first and last name). Destination servers are roughly divided into two categories: app developers vs. ad \& analytics servers. The rationale is that the application servers may need the PII to perform their functionality (\eg\xspace {\em Google Maps} clearly needs location) while third party servers do not (thus causing more of a privacy leak rather than an exposure).  For each pair of (PII type, destination type) we asked the participants to rate their comfort level of sharing that PII type with that remote  server, on a scale from 0 to 3;  where 0 represents the least concern and 3 represents maximum concern (and their willingness to pay for a privacy-preserving solution). See Fig. \ref{fig:question_10} for details.

Second, we asked participants to rate their concern in real-case scenarios of PII exposures from our dataset (Sec. \ref{sec:data}). For each packet that contained a PII in our dataset, we considered a {\bf broader context} beyond just {\em PII type} and {\em destination server type} (\ie application server or ad/analytics server). We also considered the category of the app (\eg game vs navigation), whether the PII was {\em encrypted} or sent in plain text, and the {\em  frequency} of this PII being exposed by this app category.  The rationale is that the same PII type exposed may be more or less concerning to users depending on the context. For example, location exposed by a navigation app to that app's server in an encrypted packet is probably needed for the app to function, while sending a user id to an unrelated third party (\eg advertiser) server, frequently and/or in plain text, is indeed a privacy leak.

 \begin{figure}[t!]
	\begin{center}

		\subfigure[Warm-up question with a hypothetical scenario of a particular example application -- the game {\em Roblox}.]{\hspace{6pt}\includegraphics[width=0.99\linewidth]{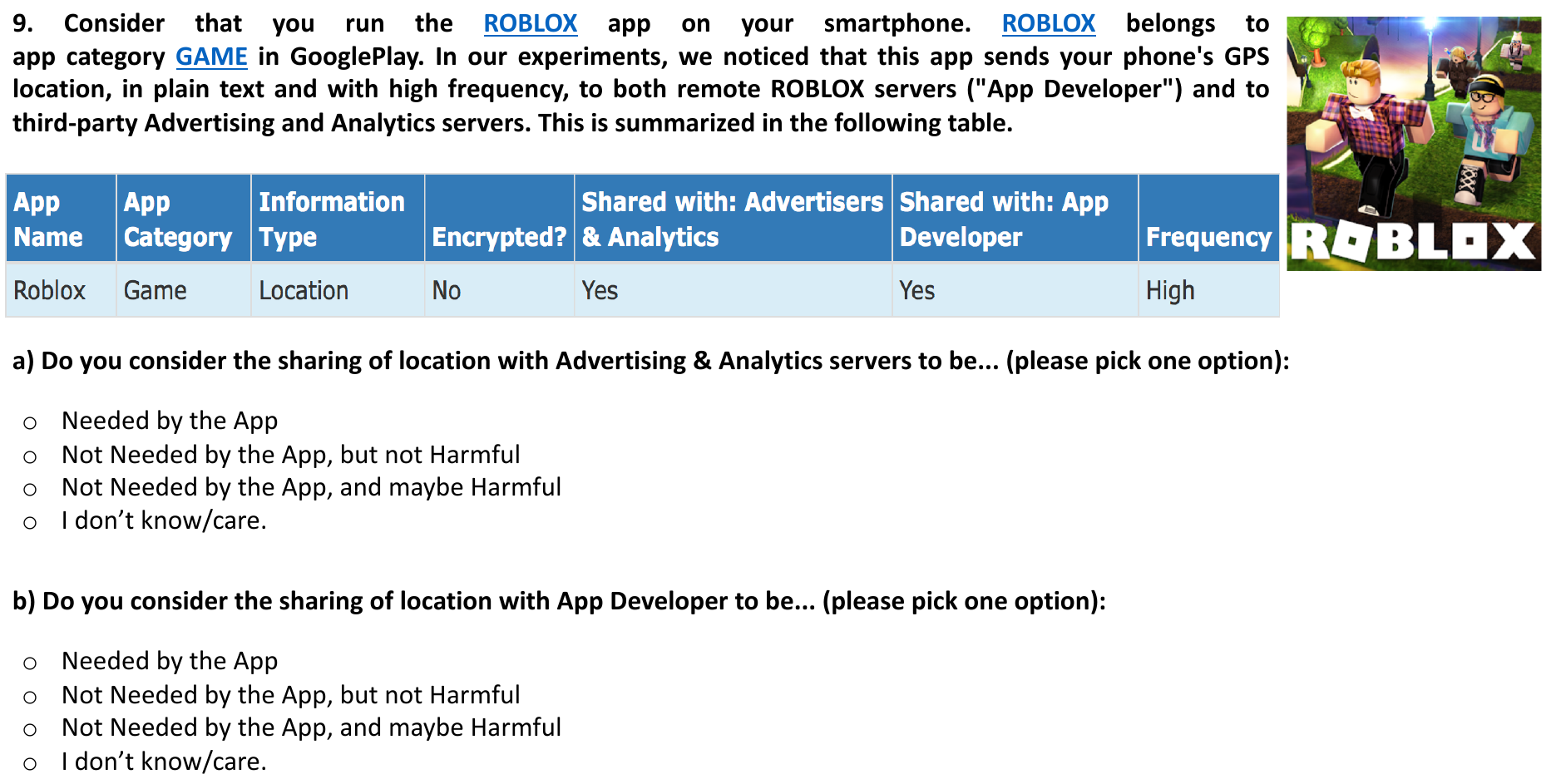}\label{fig:roblox_v2}}
		\subfigure[Entire game category with real cases of PII types exposed and their broader context (remote  server type but also app category,  encrypted/plain text, and frequency).]{\hspace{6pt}\includegraphics[width=0.99\linewidth]{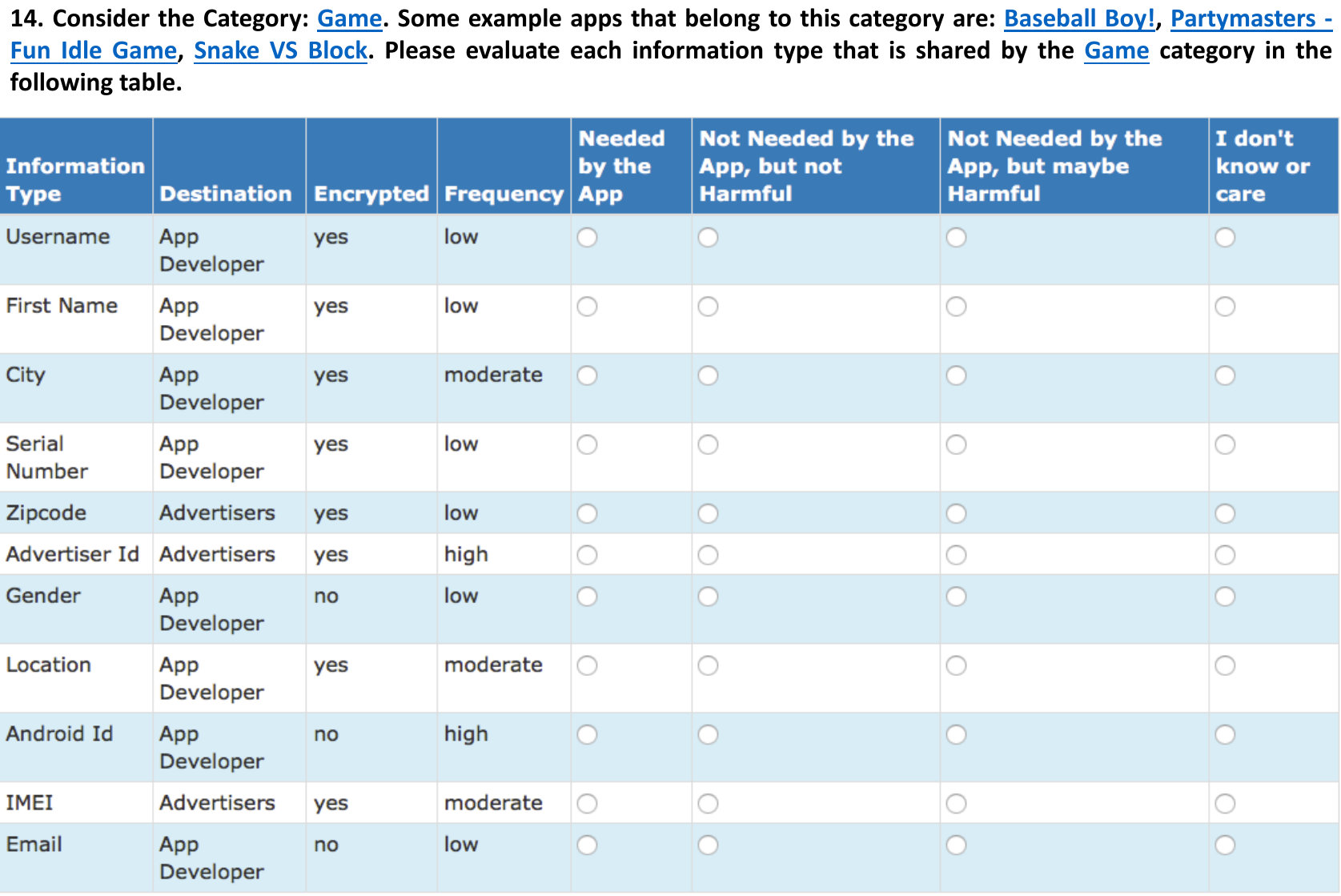}\label{fig:game_categorization}}
		%
		\caption{Task to assess user concern about {\bf privacy exposures in context.} }
		\label{fig:categorization_questions_game}
	\end{center}   
\end{figure}

A side benefit of the aforementioned categorization is that it helped reduce the number of cases to be evaluated by users. Out of 8,579 exposures total (Table \ref{tbl:summary}), 1,726 are unique when considering the application responsible, the type of PII, the destination host, and level of encryption. To further reduce the number of cases to label, we grouped the applications based on their {\em Google Play Store} \cite{googleplay} category and destination type (ad \& analytics or not). To find ad \& analytics domains, we used the {\em hpHost} \cite{hphosts} list as it was found to be the most comprehensive list for the mobile ecosystem to date \cite{vallina2018tracking}. These grouping reduced the total number of unique combinations to 256, which contained 23 unique {\em Google Play Store} categories (out of 36 total). We split these combinations into five batches of HITs, where each HIT contains five (or three for the last batch) categories of apps to be labeled.
To prepare the participants for the labeling task, we first showed a ``warm up'' question (Fig. \ref{fig:roblox_v2}) with a hypothetical scenario of the {\em Roblox} app exposing certain PII and asked them to assess the risk (Fig. \ref{fig:survey_table_categories}). Next, we asked the participants to label the exposure scenarios for each of the five categories in their HIT -- example shown in Fig. \ref{fig:game_categorization}. We also provide an example app out of each category (from our actual dataset) along with a link to the app's {\em Google Play Store} page.

\begin{figure}[t!]
	\begin{center}
		\subfigure[How much do users care about privacy?]{\hspace{6pt}\includegraphics[width=0.8\linewidth]{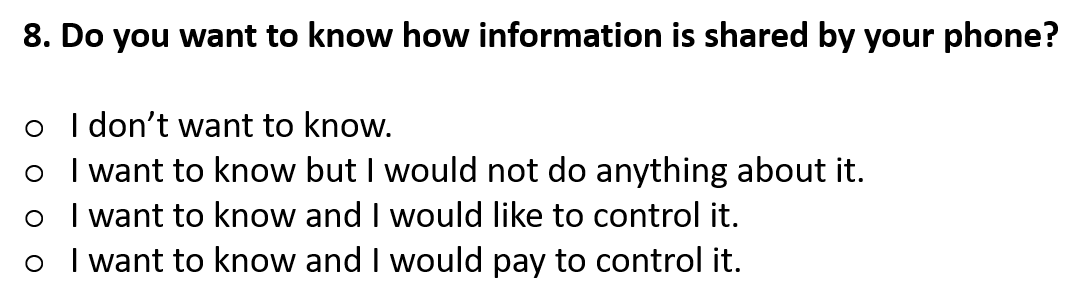}\label{fig:question_8}}
		\subfigure[How much would users pay to protect their privacy?]{\hspace{6pt}\includegraphics[width=0.8\linewidth]{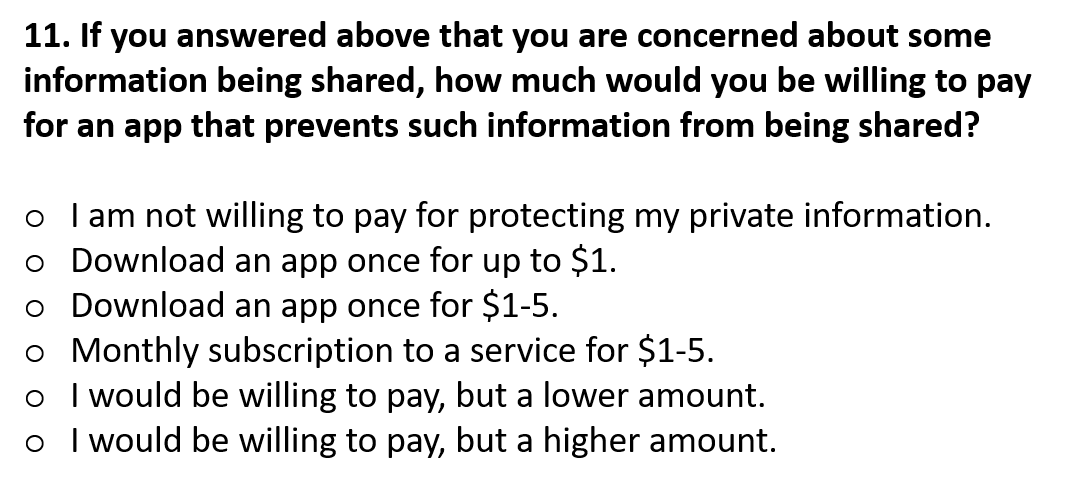}\label{fig:question_11}}
		\subfigure[Would users contribute their data to a privacy app?]{\hspace{6pt}\includegraphics[width=0.8\linewidth]{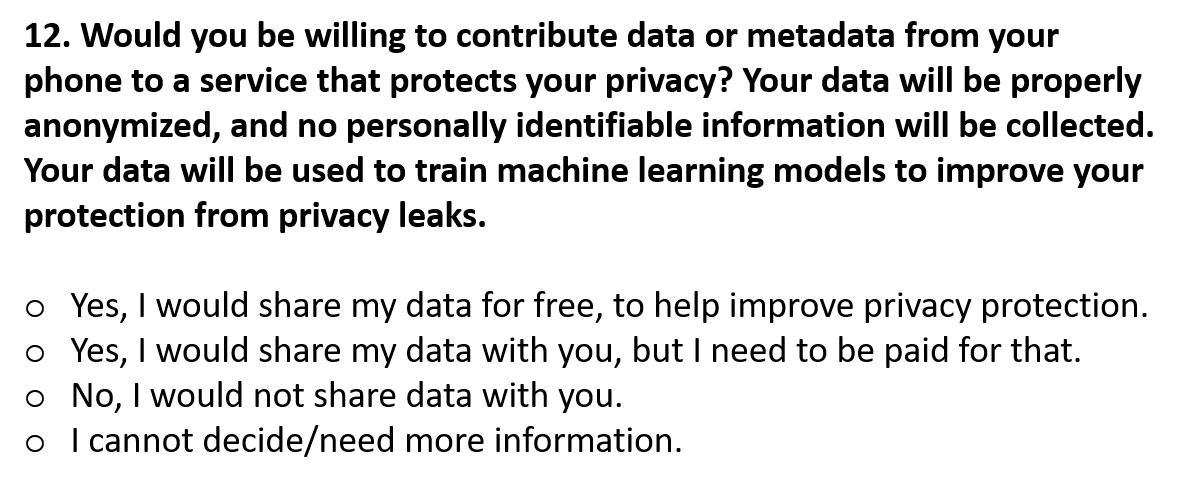}\label{fig:question_12}}
		\subfigure[Do users care more after being educated about data brokers?]{\hspace{6pt}\includegraphics[width=0.99\linewidth]{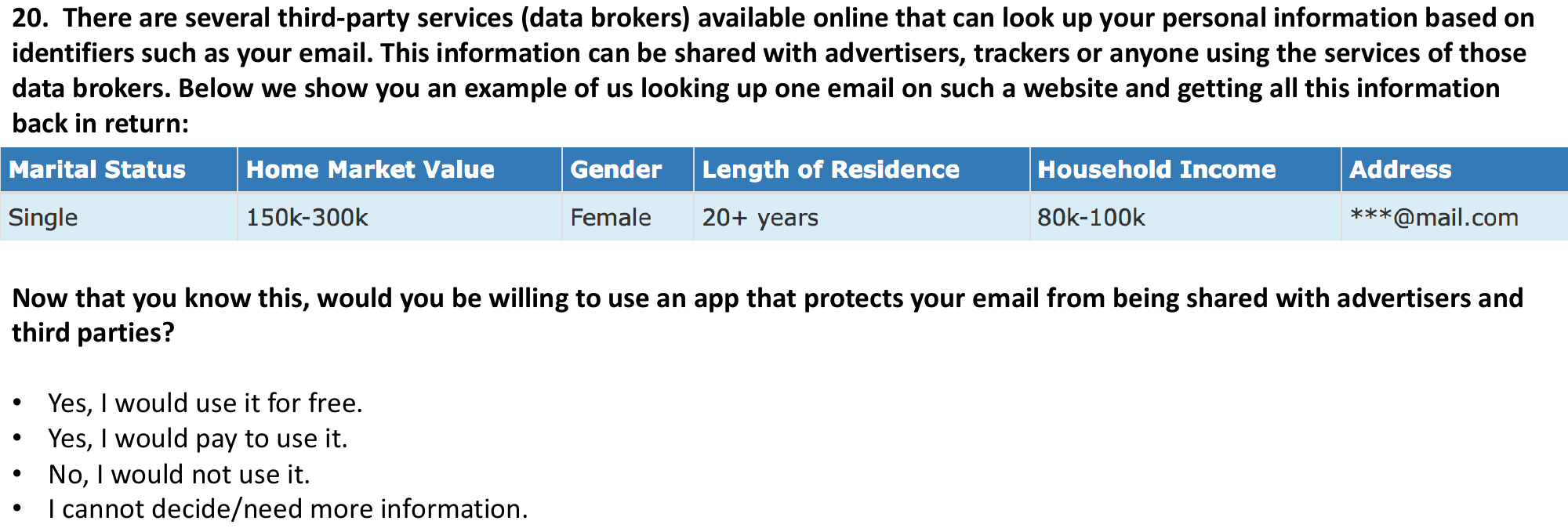}\label{fig:question_20}}
		%
		\caption{Assessing users concern and potential actions. (a) Do users care about privacy?  What they are willing to do about it: (c) share their data with a crowdsourcing system (b) use and/or pay for a privacy app). And (d) do they change attitude after being educated?} 
		\label{fig:questions_how_much_they_care}
	\end{center}   
\end{figure}

{\bf Assessing User Concern and Possible Actions.}  In order to assess participants' privacy awareness and understanding, we asked another set of questions shown in Fig. \ref{fig:questions_how_much_they_care}. First, we asked the participants how much they care about information being shared by their mobile device (Fig. \ref{fig:question_8}) and what would they do in order to better protect their information. Next, we asked if they would use an app (\eg \anteater) that can prevent privacy exposures and how much would they pay for such a service (Fig. \ref{fig:question_11}). It was our hope that the categorization questions described previously would educate users about mobile privacy and would make them more concerned towards the end of the survey. To educate them further, we showed them that a single PII in the hands of a data broker can help create or lookup a user profile and can reveal much more information (Fig. \ref{fig:question_20}); the question was based on a real scenario where we fetched data from a data broker based on a person's email address. 
We then asked them (if they would use a privacy app 
(and if they would pay for it) to protect their privacy. 
Finally, we asked them if they would contribute their mobile data to an app that crowdsources information, train machine learning models and prevents privacy leaks (\eg as in \recon \cite{recon15} and in \cite{antshield-arxiv}). 
In order to assess our hypothesis that users became more privacy-aware after the categorization questions and the data broker example, the same questions appeared twice during the survey: once in the middle of the survey and once at the end.

\subsection{User Study Results} \label{sec:study_results}

In this section, we summarize and analyze the main results of our user study. The main observation is that users seem initially confused about the severity of PII shared in different contexts, but they are interested in and are capable of being trained. Our main findings are as follows:

	\begin{itemize}
		\item Users do not seem to understand how severe it is to share certain PII types (especially device identifiers) with either Advertisers or Developers. This may be because some of these ids, such as Android ID, IMEI, IMSI, ICCID, are difficult to understand or relate to, yet they uniquely identify the device and/or the user and hence should not be shared with remote servers. 
		\item As expected, users trust Application Developers more than Advertisers \& Analytics. However, some comments stated that Developers are ``a bunch of hackers behind servers,'' indicating possible confusion.
		\item Users also seem confused about sharing PII in plain text vs. using encryption. 
		Sharing PII in plain text is a bad practice, since it exposes private information not only to the destination servers but also to anyone that is sniffing into the network.
		\item Towards the end of the study, most users seem to obtain a much better understanding of mobile privacy. For example, several comments state that users are grateful for our short tutorial in mobile privacy. They are willing to educate themselves more in the future and to adopt data transparency and privacy tools. These are encouraging results for future work in developing privacy-enhancing technologies.
	\end{itemize}

\subsubsection{Summary and Demographic Info.} 
We collected 223 responses, in total,  on our MTurk survey. Since we posted multiple HITs, each with different categorization questions (see Sec. \ref{sec:study_design}), some users completed more than one HIT, and there were a total of 151 unique participants. Two responses were discarded due to failing the attention-check questions, and one worker was discarded due to incomplete answers. This leaves a total of \numUsers valid responses, which we analyze in the rest of the section. 
The majority of the participants were between 25 and 34 years old, held a Bachelor's degree and were employed in a non-tech sector. 61.8\% and 37.7\% of the workers were Android users, and iOS users, respectively. 118 of our workers use six to ten different apps every day, 53 use between 11 and 20 apps, 41 use fewer than five apps, and only 8 use more than 20 apps.



\subsubsection{Categorization Results: Privacy Exposures in Context.}

In this subsection, we provide the results we obtained by processing user answers to the categorization questions (see Figures \ref{fig:question_10} and \ref{fig:categorization_questions_game}). We asked participants to rank the severity of PII exposures in different contexts since whether or not an exposure is considered a privacy leak depends on the context. There are four main dimensions in each exposure: (i) its destination (app developers or advertisers), (ii) category of the app responsible, (iii) level of encryption used, (iv) PII type. These dimensions play an important role in distinguishing privacy leaks from exposures. 

\begin{figure*}[t!]
	
	\subfigure[Rated by users]{\hspace{6pt}\includegraphics[width=0.2\linewidth]{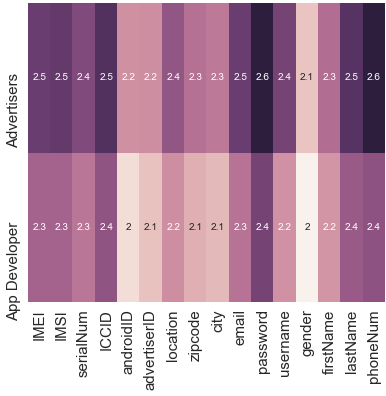}\label{fig:comfort_to_share}}
	\subfigure[Rated by us]{\hspace{6pt}\includegraphics[width=0.45\linewidth]{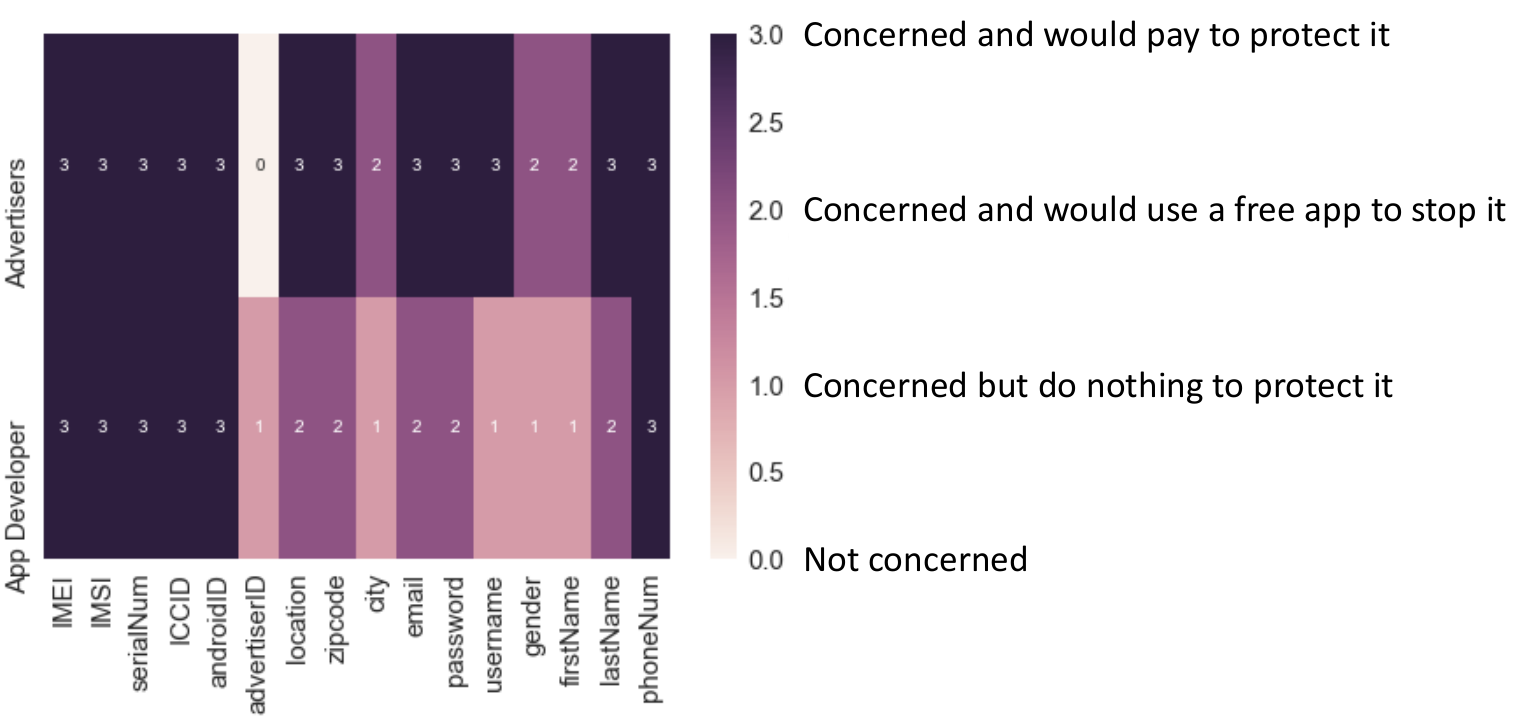}\label{fig:ideal_heatmap_comfort}}
	\caption{How comfortable are users with sharing PII with advertisers and app developers? (a) Average rating of user responses to categorization task  of Fig. \ref{fig:question_10} (b) Recommended rating by us (``experts''). }
		\label{fig:heatmap2}
	\vspace{-15pt}
\end{figure*}

Figure \ref{fig:comfort_to_share} demonstrates the results when we asked the users how concerned they are with sharing a certain information type with a particular destination (Advertisers \& Analytics or App Developer servers), as shown previously in Fig. \ref{fig:question_10}. In this task, we observe that sharing all types of PII is concerning for our participants, regardless of its destination, and they are willing to use a free app to protect them. Furthermore, they would also pay for a tool that protects their phone number and password from leaking to advertisers.
As expected, participants are more concerned over their precise longitude and latitude coordinates being shared as opposed to zipcode and city.
Overall, our participants seem to trust developers more than advertisers, which is not surprising. 

In Fig. \ref{fig:ideal_heatmap_comfort}, we also provide our own ``expert'' rating for comparison. In particular, we would like to protect device identifiers regardless of their destination, by using a paid solution if a free version is not available. In contrast, our participants chose only to protect their device identifiers with a free solution and they are not willing to pay to protect them. Moreover, they give similar rating to their location data regardless of the destination, although they should be more careful with advertisers than developers, as there are apps that need location in order to function. On the other hand, Advertiser ID should not be protected as well as the other identifiers, since this id is known by advertisers anyway. 

{\bf Understanding whether PII is needed by the app.}
We ask users to assess whether sharing a particular PII is legitimate and needed for the functionality of the app (or more broadly by apps in the same category), or if it is unnecessary and potentially harmful. For example, GoogleMaps is an app in the Navigation category and needs to share location to provide the service.

 \begin{figure*}[t!]
	\begin{center}
		%
		\includegraphics[width=0.85\linewidth]{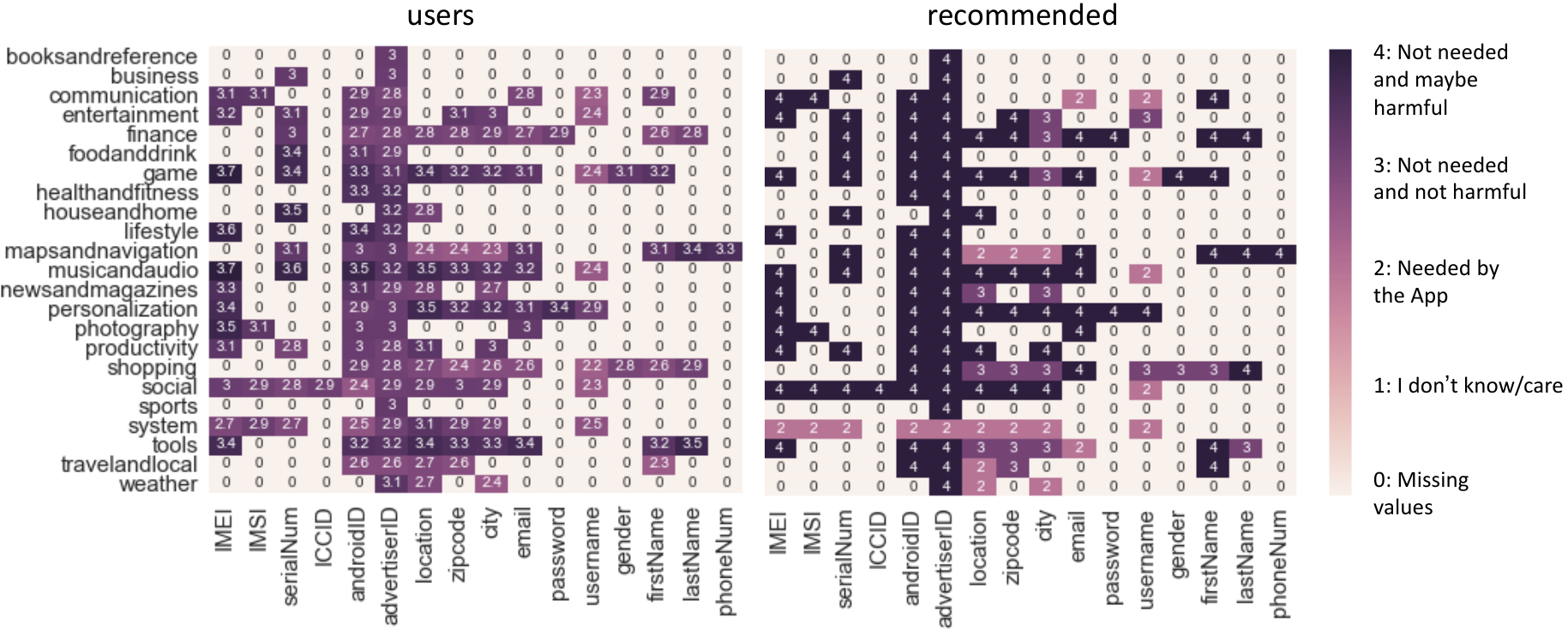}
		\caption{Heatmap severity of (PII type, app category). The darkness of the color indicates the perceived severity of the PII exposure: the darkest corresponds to 4.0 (``Not needed by the app and maybe harmful''), while the lightest indicates 1.0 (``I don't care''). Zeros represent missing values for combinations we did not have in our datasets. We compare the average ratings among users who answered that question (on the left column) vs. labeling recommended by us (on the right column).}
		\label{fig:heatmap_legitimate_non_legitimate}
		\vspace{-15pt}
	\end{center}   
\end{figure*}

Fig. \ref{fig:heatmap_legitimate_non_legitimate} presents a heatmap of the perceived severity of different PII types per application category. 
The x-axis show the different PII types and the y-axis contains the categories of apps responsible for sending that PII over the network. The values and colors represent the level of concern the participants have regarding each pair (category, PII type): ``not needed by the app and maybe harmful" (value 4 - dark color), ``not needed by the app and not harmful" (value 3), ``needed by the app" (value 2), ``don't care" (value 1 - light color). White color (or 0 value) represents missing combinations of PII and category, that were not present in our datasets. 
In order to produce the heatmap, we consider the mean values of all participants' answers. 

For comparison, we also show a heatmap labeled by us to express our ``expert'' opinion. First, location information (latitude and longitude coordinates, zipcode, and city) should be available to system, maps \& navigation, weather, and travel \& local categories only. All the other categories do not need this kind of information for their functionality and hence when apps transmit such data, it may be considered harmful. Second, user identifiers, such as email, password, username, gender, \etc may be needed in some app categories (\eg social, games, communication), but not in others (\eg photography, personalization, tools). Finally, device identifiers, such as IMEI, IMSI, Serial Number, ICCID, and Android ID should not be used by any app category, except perhaps for the System. These identifiers are unique and can be used to track users across different apps and build profiles, such as the one we showed in Fig \ref{fig:question_20}. Furthermore, Google explicitly discourages app developers from using these identifiers and asks them to instead create and use their own unique identifiers within their app \cite{googleids}. 

Comparing the two heatmaps in Fig. \ref{fig:heatmap_legitimate_non_legitimate} reveals that users are not as concerned about device identifiers as they should be: 
32.6\% of the device identifiers per app category were categorized by workers  as ``Not needed by the app and not harmful,'' although they should only be accessed by the System. However, other responses make sense: app categories that should not require any information to function are trusted the least (games, lifestyle, music \& audio, personalization, and tools), and categories such as weather and travel \& local are trusted with location information.

 \begin{figure*}[t!]
	\begin{center}
		\subfigure[Exposures in plain text: Avg user rating vs. our recommended rating]{\hspace{6pt}\includegraphics[width=0.55\linewidth]{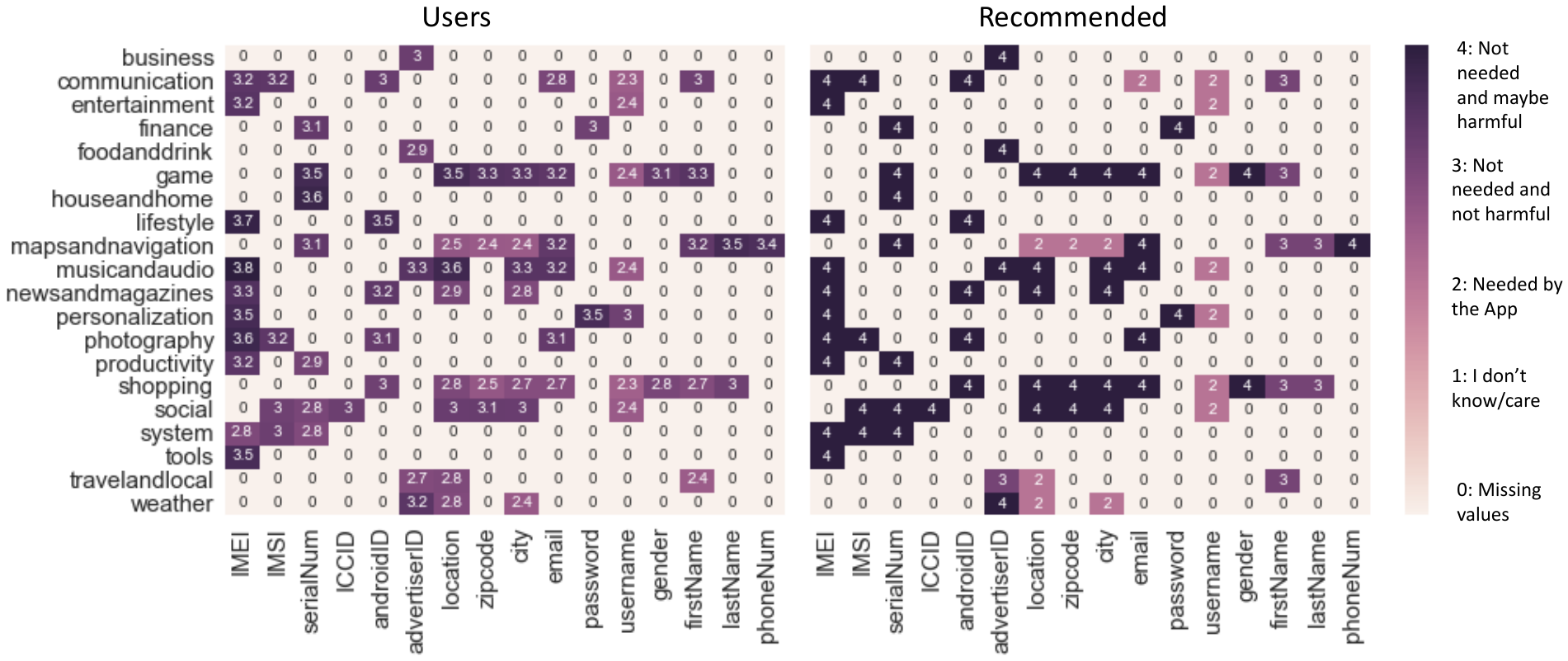}\label{fig:heatmap_unencrypted}\hspace{6pt}}
		\subfigure[Exposures in encrypted traffic: average user rating]{\hspace{6pt}\includegraphics[width=0.35\linewidth]{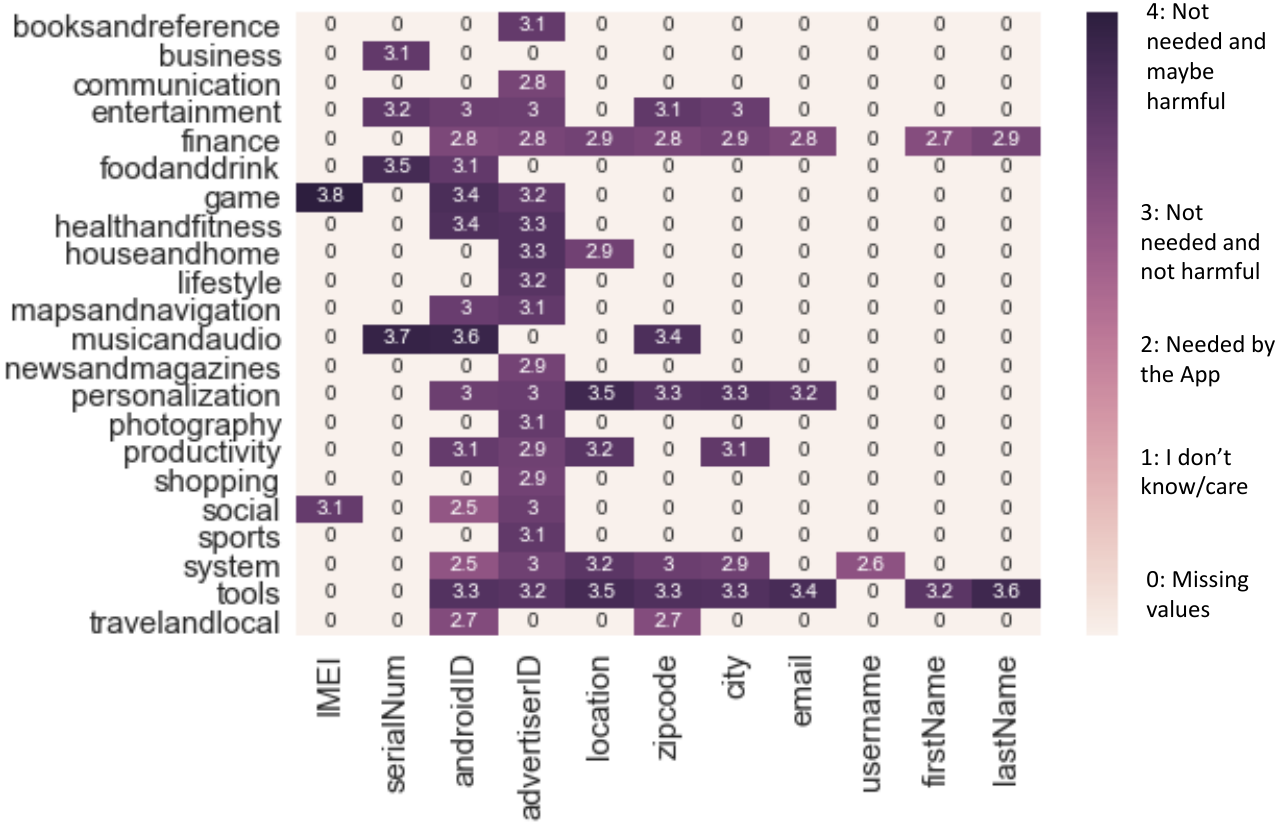}\label{fig:heatmap_encrypted}\hspace{6pt}}
		\caption{Heatmaps assessing the severity of exposures in encrypted vs. unencrypted packets, as assessed by users  vs. us (``experts'').} 
		\label{fig:heatmaps_encryption}
		\vspace{-15pt}
	\end{center}   
\end{figure*}

{\bf Understanding Encryption.} 
We further split the heatmap in  Fig. \ref{fig:heatmap_legitimate_non_legitimate} into  Fig. \ref{fig:heatmap_unencrypted} and Fig. \ref{fig:heatmap_encrypted}, depending on whether the unencrytetd packet containing the PII was unencrypted or encrypted, respectively. Our heatmaps not only reveal participants' opinions about exposures, but also showed behavioral patterns of app categories. From Fig. \ref{fig:heatmap_unencrypted}, we see that the weather category does not use encryption and sends Advertiser Id, Location and City to remote servers in plain text. Similarly, maps \& navigation category sends Serial Number, Location, Zipcode and City information in plain text, but encrypts the Android ID and Advertiser ID. Sending PII in plain text is more harmful since this traffic can be sniffed. Unfortunately, our MTurk participants did not seem to understand the implications of transmitting data in plain text.

\begin{figure}[t!]
	\begin{center}
		\subfigure[PII sent to Ad Servers.]{\hspace{6pt}\includegraphics[width=0.8\linewidth]{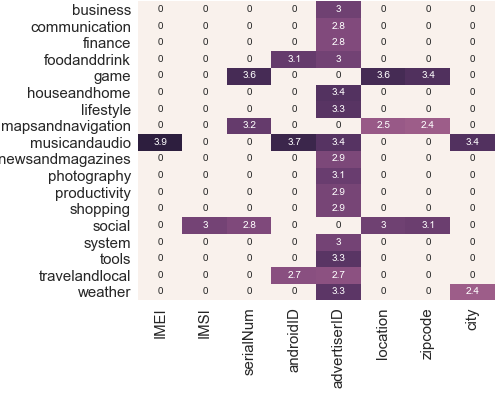}\label{fig:advertisers_destination_heatmap}\hspace{6pt}}
		\subfigure[PII sent to Developer Servers.]{\hspace{6pt}\includegraphics[width=0.99\linewidth]{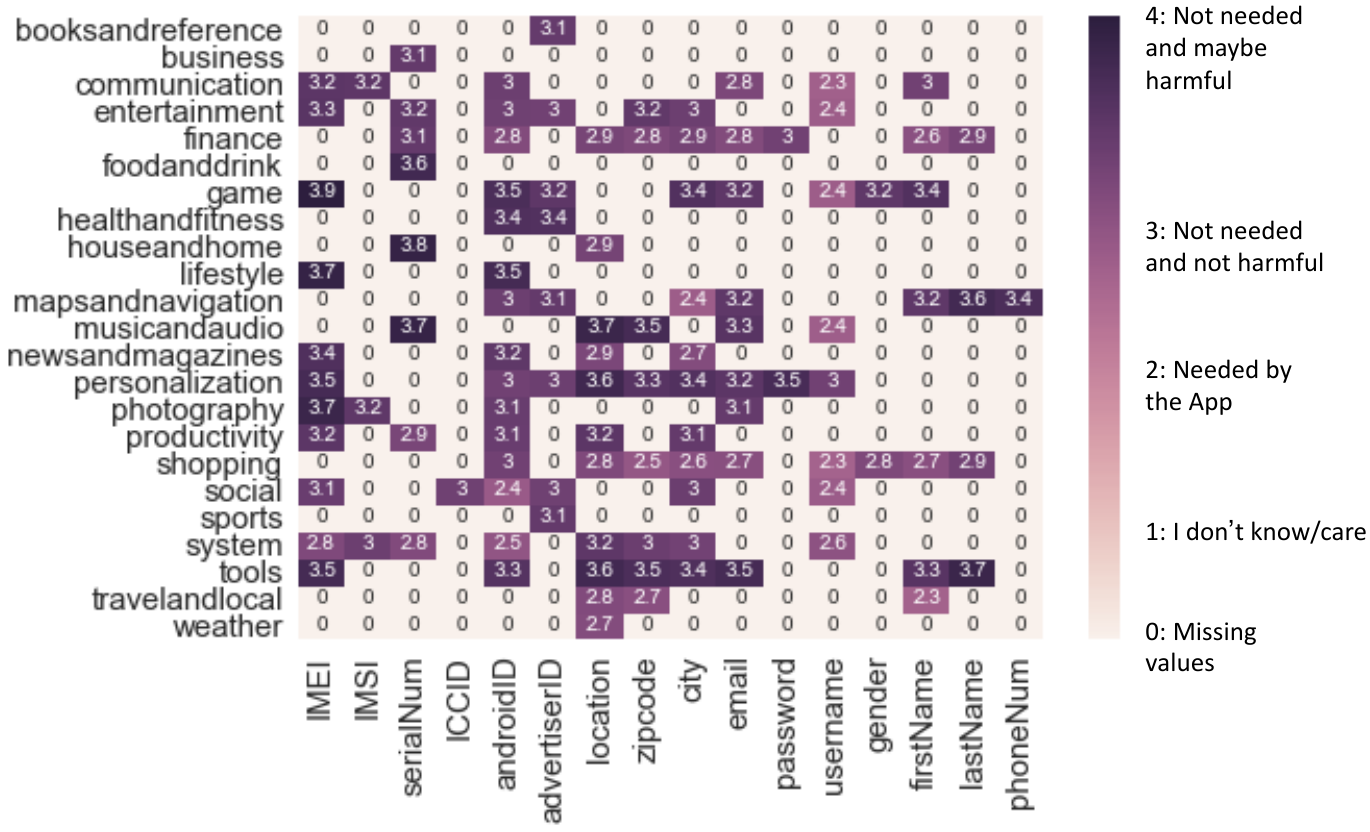}\label{fig:developer_destination_heatmap}\hspace{6pt}}
		\caption{Heatmaps assessing the severity of PII sent to different destination: Third-party (Ad Servers) vs  App Developers. Assessed by users (average user rating shown). The darker the color the more concerned the users).}
	\end{center}
\end{figure}

{\bf Understanding Destinations.} We also split our heatmaps based on whether the packet is going to App Developers (Fig. \ref{fig:developer_destination_heatmap}) or Advertiser \& Analytics Servers (Fig. \ref{fig:advertisers_destination_heatmap}). As expected, developers require more types of information in contrast to advertisers. On the other hand, advertisers should ideally only require the Advertiser ID, which is indeed fetched by almost every app category. However, most of our participants indicated that none of the PII need to be sent to advertisers for the app's functionality (only 2 out of 31 values in Fig. \ref{fig:advertisers_destination_heatmap} are below 2.5). This indicates that perhaps users don't consider ads being served a part of the app's functionality. In contrast, participants indicated more trust towards app developers (Fig. \ref{fig:developer_destination_heatmap}), which is expected as certain app categories require PII to function correctly. For example, the following pairs of PII and categories are expected: username for applications with logins (communication, games, and social), email for communication, and location information for travel \& local and maps \& navigation. However, device identifies, such as IMEI, IMSI, Serial Number and Android Id are not needed by any app category and they should not be shared with advertisers nor developers. Our participants seem to not understand this point and they labeled most of these exposures as ``Not needed and not harmful,'' while in fact these ids are most likely used for tracking. One interesting finding is that participants showed more concern over their IMEI vs. other device identifiers. This indicates that they may need more education about the other device identifiers.


\subsubsection{Towards A Solution: Tools Enhancing Data Transparency and Privacy.} 

\begin{figure}[t!]
	\begin{center}
		\subfigure[How much users care about privacy exposures.]
		{\hspace{6pt}\includegraphics[width=0.45\linewidth]{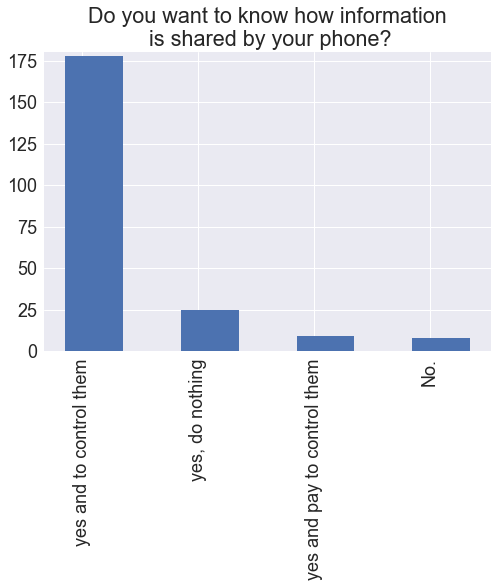}\label{fig:question_8_ans_know_about_leaks}}
		\subfigure[Monetization question at the beginning of the study.]{\hspace{6pt}\includegraphics[width=0.45\linewidth]{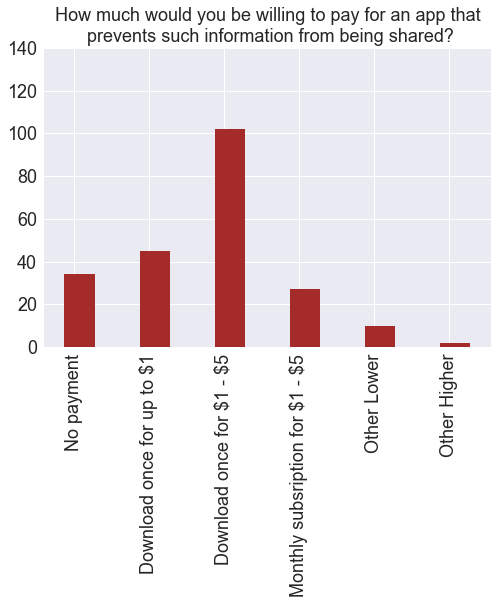}\label{fig:payment_before}}
		\subfigure[Monetization question at the end of the study.]{\hspace{6pt}\includegraphics[width=0.45\linewidth]{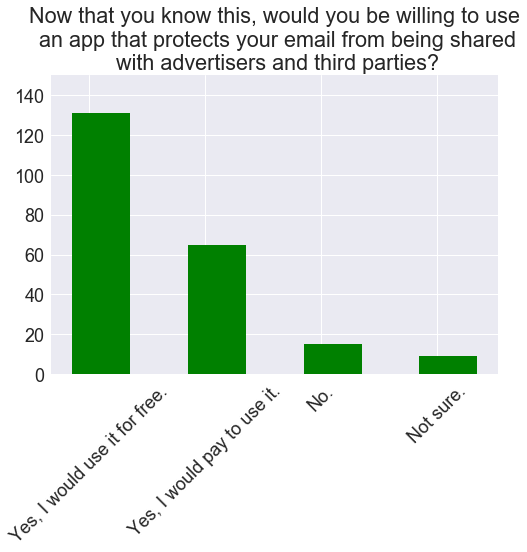}\label{fig:payment_after}}
		\subfigure[Data contribution in the beginning vs. at the end of study]{\hspace{6pt}\includegraphics[width=0.45\linewidth]{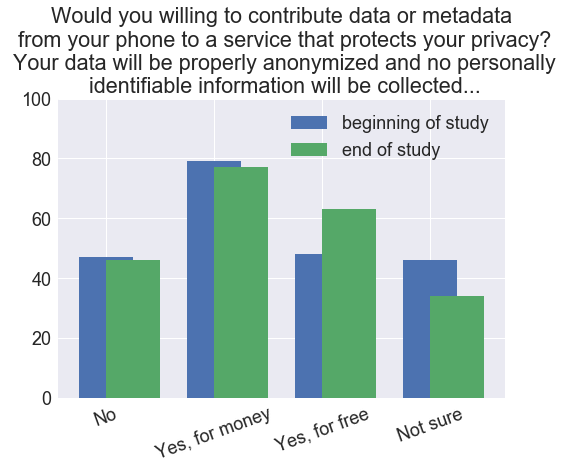}\label{fig:all_data_contribution_before_after}}
		%
		\caption{Answers to questions of Fig.  \ref{fig:questions_how_much_they_care}: how much do users care about mobile privacy and what they are willing to do to protect it? Some  questions are purposely repeated in the beginning and at the end of the study.} 
		\label{fig:answers_how_much_they_care}
	\end{center}   
\end{figure}

Fig. \ref{fig:question_8_ans_know_about_leaks} shows the distribution of answers to the questions of Fig. \ref{fig:questions_how_much_they_care}, which essentially ask how much users care about privacy exposures and what they are willing to do about it. The overwhelming majority of users would like more control over their information being shared, if a free option is available. Fig. \ref{fig:payment_before} demonstrates what payment options users would prefer for a privacy app: the majority would prefer a one-time fee between \$1 and \$5, than a subscription model. These results show that users are not only interested in using privacy-preserving tools  but are also willing to pay for them.
Towards the end of the study, after being educated about the extent and severity of  sharing PII, more users were willing to pay for a privacy app, \eg compare Fig. \ref{fig:question_8_ans_know_about_leaks} to Fig. \ref{fig:payment_after}. This demonstrates that although users may originally not be aware or understand the risks of privacy exposures, they become more weary after being educated about  the occurrence and risks of sharing PIIs (especially after learning the power of data brokers). 
Finally, Fig. \ref{fig:all_data_contribution_before_after} shows the user's willingness to contribute data to a privacy-preserving tools that crowdsource information (such as \haystack, \recon, \anteater) before and after completing our survey. Once again, completing the study made users more open towards using and helping a privacy-preserving app. 

\subsubsection{User Comments.}
\begin{figure}[t!]
	\begin{center}
		\subfigure{\hspace{6pt}\includegraphics[width=8.5cm]{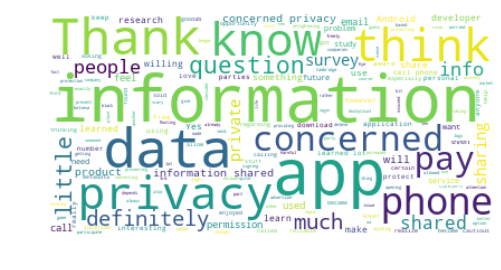}\label{fig:all_comments}}
		%
		\vspace{-10pt}
		\caption{Overview of main keywords extracted from 99 MTurk workers' comments} 
		\label{fig:comments}
	\end{center}   
\end{figure}

At the end of the study, we asked participants for their comments and suggestions; 99 out of \numUsers users provided their feedback. In Fig. \ref{fig:comments}, we summarize the comments of all workers in the form of a wordcloud.
Overall, our \mturk workers seemed satisfied with the survey and stated that it was an educational experience. 
Most of the participants are thankful for our short tutorial on mobile privacy, as they gained more knowledge about privacy and how different apps share their information with various destinations. At the end of the survey, they seem more concerned about privacy and interested in a solution, including a privacy app free or paid; see Fig. \ref{fig:answers_how_much_they_care}.       
Several participants mentioned the recent Facebook and Cambridge Analytica scandal and wondered if our user survey was inspired by it (it was not!). 
Below, we provide a sample of representative user comments, out of 99 total comments, grouped by two recurrent topics. 

\begin{quote}

\item {\bf They learned a lot:}
\item \textit{``I liked this. Was this all due to what is going on with Facebook and other privacy	concerns?''}
\item \textit{``I enjoyed taking this survey.  I would not share my information with anyone that would use the information in anyway that would be damaging to my life.  Right now I am getting phone calls I don't want.  I would pay to have those stopped.''}
\item \textit{``I definitely became more concerned about how much data is taken. You don't
	realize how much of your personal data is sent to advertisers and it makes you
	more weary of downloading and using certain apps.''}
\item \textit{``I thought it was quite enlightening. I will certainly be paying more attention to what and how an app uses my data in the future.''}
\item \textit{``I definitely became more concerned about my privacy. Thank you for the wake up call.''}
\item \textit{``Good Study. Made me really think about how much of everybody's information is really out there.''}
\item \textit{``I am surprised to learn how much information could be linked to my email address.  It is like telemarketing on steroids and I would be willing to keep that info private  I think we should always have the option to keep our private info private.''}
\item \textit{``I have become a little more concerned about my privacy as some of this was new information to me. I will definitely be doing more research on this subject in the near future. Thank you for the informative study! Everything was very clear and straight forward, I appreciate the opportunity to participate.''}
\newline
\item {\bf They are interested in using a privacy-enhancing solution:}
\item \textit{``I did not know all of that and if you developed a product I want it.''}
\item \textit{``Yes - I have been concerned, but it has been difficult to know where to start with securing my data.  I live in the US, so there is no GDPR to protect me, however I feel I can review application requests for information in a more informed manner.  I would also love to find the service you mentioned about protecting privacy of data.''}
\item \textit{``Thank you. I learned a lot from this study. I hope you are able to develop something that can help protect consumers and still allow developers to flurish.''}
\item \textit{``I would love a reliable app that limited/reduced data collection and increased privacy. Problem is that it's not possible to know which privacy apps are reliable and effective, while still allowing a service or app to be used. Even a reliable, mainstream email alternative would be good!''}


\end{quote}

\section{Conclusion \& Future Directions} \label{sec:conclusion}

We provided a combination of a measurement and a user study of actual PII exposed by mobile apps.  We also defined and analyzed the context (PII type, destination domain, app category/functionality, background/foreground, use of encryption vs. plain text)  where these PII exposures occur, and we distinguished between PII exposure and PII leak (which is more likely to be harmful) depending on the context. In the measurement study, we collected and analyzed a new richer dataset, which reveals  interesting PII exposures and patterns, some of which were previously unknown. Preliminary graph analysis revealed interesting patterns of apps and domains colluding to expose private information.  In the user study, we compiled the large amount of information from the measurement study into a smaller number of categories (contexts) and we asked users  to assess the privacy exposures in the actual context they appeared, w.r.t. their legitimacy and risk they pose. Most users were initially unaware of the severity and potential implications of PII exposures: they could not identify the most critical PII or context (\eg that it may be ok to share information with the application server instead of third parties, such as advertises and trackers). However, they seemed to appreciate the information they got through the study, which made them more willing to adopt privacy enhancing tools. Our analysis combines the scale and coverage of the network-based measurement study with the fine-granularity user input assessing privacy exposures in context. 

There are several directions for future work, building on the observations of this paper. First, the behavioral analysis of PII leaks at the end of Section \ref{sec:data} revealed  similarities in the way apps and destination domains extract PII from mobile devices. Those observations can be further exploited to design machine learning approaches that can detect packets with potential privacy exposures \cite{spawc_shuba, bakopoulou2019federated}, further inspect them and eventually prevent an actual leak (\eg by using real-time  tools like \anteater to block a packet or obfuscate a PII). Second, the user study showed that users are interested in and are capable of  being educated about their data, and they want to adopt better privacy practices and new tools (such as \anteater, \recon or \haystack) to enhance data transparency and privacy control.

\section{Acknowledgments}\label{sec:acknowledgements}

We would like to thank Milad Asgari Mehrabadi for providing figures 2 and 3 for this paper.

This work has been supported by NSF Awards 1649372, 1815666, 1900654 and by a UCI Proof-of-Concept Award in 2017. E. Bakopoulou has also been supported by the Broadcom Foundation and Henry Samueli fellowships. A. Shuba  has also been supported by an ARCS Fellowship.

\bibliographystyle{plain}
\bibliography{allrefs_sorted}

\end{document}